\documentclass{custom} 

\title[An OT approach to estimating causal effects via nonlinear DiD]{An optimal transport approach to estimating causal effects via nonlinear difference-in-differences}

\usepackage{amsmath}
\usepackage{amssymb}
\usepackage{dsfont}
\usepackage{natbib}
\usepackage{bbm}
\usepackage{color}
\usepackage{enumitem}
\usepackage{graphicx}
\usepackage{mathtools}
\usepackage{microtype}
\usepackage{tikz}
\usepackage{times}
\usepackage[Symbolsmallscale]{upgreek}
\usepackage{thm-restate} 
\usepackage{nicefrac} 

\makeatletter
 \let\Ginclude@graphics\@org@Ginclude@graphics 
\makeatother

\coltauthor{\Name{William Torous}\thanks{WT was supported through MIT's UROP program with funding from the Lord Foundation and the Paul E. Gray UROP Fund} \Email{wtorous@berkeley.edu}\\
 \addr University of California, Berkeley\\
 \Name{Florian Gunsilius}\thanks{FG is supported by a MITRE faculty research award} \Email{ffg@umich.edu}\\
 \addr University of Michigan\\
 \Name{Philippe Rigollet}\thanks{PR is supported by NSF grants IIS-1838071, DMS-2022448, and CCF-2106377} \Email{rigollet@math.mit.edu}\\
 \addr Massachusetts Institute of Technology}

\begin{document}

\maketitle

\begin{abstract}%
We propose a nonlinear difference-in-differences method to estimate multivariate counterfactual distributions in classical treatment and control study designs with observational data.  Our approach  sheds a new light on existing approaches like the changes-in-changes- and the classical semiparametric difference-in-differences estimator and generalizes them to settings with multivariate heterogeneity in the outcomes. The main benefit of this extension is that it allows for arbitrary dependence and heterogeneity in the joint outcomes. We demonstrate its utility both on synthetic and real data. In particular, we revisit the classical Card \& Krueger dataset, examining the effect of a minimum wage increase on employment in fast food restaurants; a reanalysis with our method reveals that restaurants tend to substitute full-time- with part-time labor after a minimum wage increase at a faster pace.\footnote{A  previous version of this work was entitled "An optimal transport approach to causal inference."}
\end{abstract}


\section{Introduction}

Difference-in-differences estimators are among the most widely-used approaches to estimate causal effects of discrete treatment interventions in observational data \citep{abadie2005semiparametric, lechner2011did, imbens2015causal, roth2022s}. The classical setting we considered in this article is one in which the researcher observes the outcomes of interest in a group of treated units and a corresponding control group of untreated units over time, i.e.~a binary treatment setting.\footnote{The proposed method can straightforwardly be extended to finitely many treatments. We do not address the question of continuous treatments in this article.} At one fixed point in time the treatment group undergoes an intervention and stays treated thereafter. The fundamental problem consists of isolating the causal effect of this intervention from the existing \emph{natural trend} the treatment group undergoes irrespectively. The method of difference-in-differences achieves this under the assumption that the units in the control group have the same fundamental trend as the treated units would have without intervention. Under this assumption, the difference of the average outcome of the treated group between pre- and post-intervention periods net the difference of the control groups average outcomes pre-and post intervention manages to isolate the true average causal effect of the intervention.

The arguably two most fundamental methods for estimating causal effects in this setting are the classical method of Difference-in-Differences (DiD) \citep{abadie2005semiparametric, lechner2011did, wing2018designing} as described above and its generalization beyond average and aggregate effects, the Changes-in-Changes (CiC) estimator \citep{athey2006identification}.

DiD is a linear approach designed to estimate average (or aggregate) causal effects \citep{abadie2005semiparametric, heckman1990varieties} over all units within a group. This method can be proved to correctly identify counterfactuals under the ``parallel trends'' assumption~\citep{abadie2005semiparametric, roth2020parallel, ryan2019now} under which the \emph{average} natural trend is assumed to be the same across both the control and treatment group. This idea has been used in many areas of science where capturing the average treatment effect is sufficient to formulate informative causal conclusions. Recent applications include quantifying the effect of public health measures in response to COVID-19 \citep{bacon2020using} and estimating how irrigation farmers adapt their watering to fixed usage limits \citep{drysdale2018adaptation}.

The classical linear DiD estimator only provides estimates of average and aggregate treatment effects. This can be limiting in many settings where treatment heterogeneity is important, i.e.,~where different units can react differently to the same level of intervention. The CiC estimator \citep{athey2006identification} addresses this issue for univariate outcomes. It extends the fundamental idea of the DiD estimator by estimating changes in the entire probability distribution of the respective units within a group. This permits the estimation of the entire counterfactual law of the treated units had they not received the treatment, hence allowing for a general form of heterogeneity in their response to treatment. The cost is a slightly stricter assumption on the evolution of the unobservable distribution as we specify below.

In its current form the CiC estimator is only applicable in settings with univariate outcomes, as it relies heavily on the definition of quantile functions. This can be restrictive for modern applications where the outcomes of interest are often multivariate. Examples range from A/B testing in digital marketing campaigns where outcomes are a combination of features such as click-through rate, time-per-page, etc., to measuring intervention effects such as gene knock down on a population of cells measured in high-dimensional gene space. In principle, the CiC estimator may be extended to higher dimensions through tensorization, which estimates treatment effects independently for each coordinate, but this solution fails to capture correlations that are often important in causal discovery. We demonstrate a simple linear setting with synthetic data where this is the case in Section \ref{sec:motivating_example} below. \\

\noindent {\bf Our contribution.} We recast the the CiC estimator using tools from the theory of optimal transportation. This perspective allows us to extend this methodology to handle multivariate observables after introducing novel structural assumptions on the natural trend underlying both treatment and control groups. In particular, we note that the causal model proposed in~\citet{athey2006identification} readily implies that both populations, treatment and control, evolve between pre- and post-treatment periods via optimal transport maps (Theorem~\ref{thm:main}); our methodology relies on a natural higher-dimensional extension of this via optimal transport theory. In fact, we show that if the corresponding functions mapping the unobservables to the potential outcomes are \emph{cyclically comonotone} then the counterfactual distributions and causal effects can be estimated from data. In particular, this allows for the unobservable variables to be of the same dimension as the dimension of the outcome variables and not just univariate.

The counterfactual distributions can be estimated consistently from data using recent results from statistical optimal transport and implemented efficiently using recent advances in computational optimal transport~\citep{peyre2019computational}. We demonstrate the benefits of this extension by comparing it to the simple multivariate extension of the CiC estimator both on artificial and real data. As illustrated in Section \ref{sec:motivating_example}, a tensorized version of CiC can be inconsistent and even estimate opposite correlation structures in the multivariate distribution of counterfactual outcomes. We also revisit the classical dataset of \citet{card1994minimum} that has sparked an intense debate about the effects of raising the minimum wage on employment. Our ability to jointly handle the number of part-time and full-time workers as a bivariate outcome reveals an interesting substitution effect from the perspective of labor economics: when the minimum wage is raised, restaurants increase full-time labor and decrease part-time labor suggesting a substitution effect.

\section{The Causal Model Behind the Changes-in-Changes Estimator}\label{sec:classical}

\subsection{The changes-in-changes model}
We recall in this section the general causal model on which the classical difference-in-differences and changes-in-changes estimators are built, and which our model extends. See Figure~\ref{fig:connectionpic} for illustration.

\begin{figure}[h!t]
\centering
\begin{tikzpicture}
\draw node[left] at (-0.4,3) {$\textbf{Control}$};
\draw node[right] at (3,3) {$\textbf{Treatment}$};

\draw node[left] at (-3.5,2) {$\textbf{t=1}$};
\draw node[left] at (-3.5,-1) {$\textbf{t=0}$};

\draw node[left] at (-1.7,2.2) {$\color{magenta}\boxed{\Prob_{Y_{C,1}}}$};
\draw node at (-0.1,0.5) {$\nu$};
\draw node at (-0.1,-1) {$\nu$};
\draw node[left] at (-1.7,-1.1) {$\color{magenta}\boxed{\Prob_{Y_{C,0}}}$};

\draw[->] (-0.25,0.67) to (-1.7,1.9);
\draw[->, dashed] (-0.1,-0.75) to (-0.1,0.25);
\draw[->] (-0.3,-1) to (-1.7,-1);
\draw[->,thick, orange] (-2,-0.75) to (-2,1.75);

\draw node[left] at (-2,0.5) {$\color{orange} \sd $};
\draw node[above] at (-0.9,-1) {$h_0$};
\draw node at (-1,0.95) {$h_1$};

\draw node[left] at (2.2,-1.1) {$\color{magenta}\boxed{\Prob_{Y_{T,0}}}$};
\draw node[left] at (2.2,2) {$\Prob_{Y_{T,1}^\dagger}$};
\draw node at (4,-1) {$\nu^\star$};
\draw node at (4,0.5) {$\nu^\star$};
\draw node[right] at (5.8,2.1) {$\color{magenta}\boxed{\Prob_{Y_{T,1}}}$};

\draw[->] (3.7,-1) to (2.2,-1);
\draw[->, dashed] (3.9,-0.75) to (3.9,0.2);
\draw[->] (3.75,0.6) to (2.2,1.8);
\draw[->] (4.25,0.6) to (5.8,1.8);
\draw[->,thick, orange] (2,-0.75) to (2,1.65);
\draw[->,thick, blue] (5.8,2) to (2.2,2);

\draw node[left] at (2,0.5) {$\color{orange} \sd $};
\draw node[above] at (3,-1) {$h_0$};
\draw node at (3,0.9) {$h_1$};
\draw node at (5.1,0.9) {$h_1^\star$};
\draw node at (4,1.75) {$\color{blue} \sT$};
\end{tikzpicture}
\caption{Illustration of various maps in the space of measures. An arrow indicates a pushforward map between two measures; for example $\Prob_{Y_{C,1}}=\sd_{\#} \Prob_{Y_{C,0}}$. The maps $h_j$ are the ``production functions'' linking the unobservables $\nu$ and $\nu^*$ to the potential outcomes. A dashed arrow indicates a map from a measure to itself. $\Prob_{Y_{T,1}^\dagger}$ is the counterfactual outcome measure of the treated units had they not received treatment. {\color{orange} $\sd$} is the natural trend map and {\color{blue} $\sT$} is map from an observed outcome to its counterfactual. The {\color{magenta} observable data} is drawn from the four boxed measures.}
\label{fig:connectionpic}
\end{figure}

We first formalize a stochastic model for an experimental design in which two groups are measured before and after some intervention. Sampled from a larger population, each unit $i$ in a realized experiment is characterized by four vector potential outcomes, their level of treatment received, and a set of time points at which potential outcomes were observed. This model includes randomized control trials with non-compliance and general observational studies. Adapting the notation of \citet{athey2006identification}, we define an indicator random variable $G_i$ for a unit's adopted treatment arm as well as random vectors $Y_{i;C,0}$ and $Y_{i;C,1}$ to model unit $i$'s potential control group pre- and post-intervention outcomes observable when $G_i = 0$; likewise $Y_{i;T,0}$ and $Y_{i;T,1}$ are the unit's observable potential outcomes when $G_i = 1.$ These potential outcomes are assumed to lie in a subspace of $\R^d,$ and each unit has indicator random variables $T_{i,0}$ and $T_{i,1}$ denoting whether an outcome was observed in each study period.

We assume that each potential outcome vector is generated by a deterministic function of a latent random vector also in $\R^d.$ While any given unit's latent vector may change over time, the population distribution of continuous latent variables is time-invariant. Furthermore, these latent distributions can differ arbitrarily between control and treatment arms. We denote this latent distribution $\nu$ for controls and $\nu^\star$ for treated. Therefore we write for control units the latent random vectors $U_{i,t} \stackrel{d}{\sim} \nu, t=0,1$ and treated $U_{i,t} \stackrel{d}{\sim} \nu^\star$. We can think of $U_{i,t}$ as capturing all of the intrinsic and unobservable characteristics of unit $i$, such as an employee's skills and motivation, which influence the outcome of interest at time $t$, for instance their salary. This assumption may be less appropriate in settings where external shocks could change the latent distributions over time. For instance in a study measuring individuals' opinion of the economy, a latent distribution of intrinsic perception reasonably shifts over time. Our model can accommodate exogenous noise through the assumption that latent $U_{i,t}$ is the additive perturbation of a unit's time-invariant baseline latent state $\mu_i$ by a stationary noise process $\varepsilon_{i,t}$. Correlation between these two terms is allowed as long as the distribution of realized latent variables is time-invariant within each treatment arm. These latent variable assumptions are general enough to include many forms of mechanistic structure researchers may know about their population.

As Figure 1 illustrates, three global \textit{production functions} $h_0$, $h_1$, and $h_1^\star$ characterize the evolution of both treatment arms over time. At $t=0$ the treatment arms are assumed to have identical factors influencing the translation of a latent variable to observed potential outcome and thus share $h_0$. For controls $h_1$ describes any changes due the natural trend over evolving time. For treated units, $h_1^\star$ captures both drift over time and the causal effect of interest. If $h_1 \circ (h_0^{-1})$ exists, it describes exactly how any given control group unit evolves over time with a fixed latent state. We can therefore obtain a counterfactual outcome distribution with this function under an assumption that the treatment arm units would have evolved in the absence of treatment just as controls do. It is notable that this proposed model does not rely on additional information from observed covariates. As such, it does not require classical unconfoundedness assumptions. However, the question about the assignment mechanism in these difference-in-differences setups is a delicate one, see for instance \citet{ghanem2022selection, marx2022parallel}.

As units in the same treatment arm share a latent variable distribution, we define random vectors with the population distribution such that $Y_{i;C,0} \stackrel{d}{\sim} Y_{C,0}$ for all units such that $G_i = 0$. The random vectors $Y_{C,1}$, $Y_{T,0}$, and $Y_{T,1}$ are defined similarly for observations adopting the appropriate treatment arms. Each population-distributed random variable of potential outcomes induces a measure or \emph{law}, which we will utilize in application of results from optimal transport theory. We adopt the notation $\Prob_{Y_{G,T}}$ for this measure. The action of each production function can be described as a \emph{pushforward} of measure. For example we write $\Prob_{Y_{C,0}} = h_{0\#} \nu$ and $\Prob_{Y_{T,1}} = h_{1\#}^\star \nu^\star$. We make an assumption that both $\Prob_{Y_{C,0}}$ and $\Prob_{Y_{T,0}}$ are absolutely continuous with respect to Lebesgue measure. In Section \ref{sec:CIC_OT} we discuss possible changes to our assumptions which would allow our methodology to handle discrete outcomes at the cost of point identification of the counterfactual distribution. We note that the discrete case in general does not allow for point-identification, which is already evident in the univariate setting, where \citet{athey2006identification} provide bounds on discrete counterfactuals. Our optimal transport framework illuminates this issue even further.

Additionally assuming that $h_0$ is invertible, we have $\nu=(h_0^{-1})_{\#} Y_{C,0}$ so that
\begin{equation}
    \label{eq:Y0toY1}
    \Prob_{Y_{C,1}} = (h_1\circ h_0^{-1})_{\#} \ \Prob_{Y_{C,0}}=:\sd_{\#} \Prob_{Y_{C,0}}.
\end{equation}

In the sequel we will propose a class of production functions such that the multivariate counterfactual distribution for the treated, distributed as a random variable ${Y}_{T,1}^\dagger$, can be identified as
\begin{equation}
    \label{eq:counterfactual_pushforward}
    \sd_{\#} \Prob_{Y_{T,0}} := \Prob_{Y_{T,1}^\dagger}.
\end{equation}

A consistent estimator for $\sd$ and hence this counterfactual pushforward is presented utilizing optimal transport. Figure \ref{fig:connectionpic} shows that the production function $h_1^\star$, which generates post-intervention outcomes for the treated, does not need to be specified to recover the map $\sd$ and its counterfactual pushforward $Y_{T,1}^{\dagger}$. As discussed in Section \ref{sec:identification}, further assumptions on $h_1^\star$ enable the treatment effect map $\sT$ to be consistently estimated and allow for the estimation of non-linear treatment effects.

\subsection{Modeling the control group time trend}
\label{sec:natural_drift}

Given two measures $\Prob_{Y_{C,0}}, \Prob_{Y_{C,1}}$ which can be estimated from control group data pre- and post-intervention, there exist infinitely many maps $\sd$ such that $\Prob_{Y_{C,1}} = \sd_{\#} \Prob_{Y_{C,0}}$ \eqref{eq:Y0toY1} holds. To identify a unique such map $\sd$, one must make additional assumptions on the causal model.

Recall that the source of randomness in the proposed causal model are latent random vectors $U_{i,t}$ which can be thought of as summarizing a unit's intrinsic characteristics. If we not only assume that the distribution $\nu$ of control group latent vectors is time-invariant but further that the latent variables are constant over time (i.e., $U_{i,0}=U_{i,1}$ for all control units $i$), then we have $Y_{i;C,1}=\sd(Y_{i;C,0})$; this map can be estimated via multivariate regression from independent copies of the pair $(Y_{C,0}, Y_{C,1})$. In many practical applications, pairs of potential outcomes $(Y_{i;C,0}, Y_{i;C,1})$ are observed, but in general the assumption that $U_{i,0}=U_{i,1}$ for all control units is more tenuous. Moreover, this setup leaves out an important class of problems where the units in the control group can differ between $t=0$ and $t=1$.  A canonical example arises in genomics where in single-cell RNA-Seq data each unit corresponds to a cell. Measuring the gene expression levels of a cell is a destructive process, and as a result a given cell may be only measured once. Therefore, coupled potential outcomes of the form $(Y_{i;C,0}, Y_{i;T,1})$ are not available. Instead, one observes independent copies of $Y_{C,0}$ as well as independent copies of $Y_{C,1}$. While difficult, this problem can be solved at the cost of additional assumptions on the natural trend $\sd$. This problem is sometimes referred to as \emph{uncoupled regression} and it arises in various applications, far beyond the genomics example described above~\citep[see, e.g.,][]{RigWee19, balabdaoui2021unlinked, slawski2022permuted}. 

In a setting with uncoupled scalar data, the map $\sd$ is unique when it is monotone with known monotonicity (either increasing or decreasing). The CiC estimator enforces this by assuming both production functions, and therefore $h_1 \circ h_0^{-1}$, are monotone increasing. In fact, the unique monotone increasing function $\sd$ such that $\Prob_{Y_{C,1}}=\sd_\# \Prob_{Y_{C,0}}$ is given by $F_1^{-1}\circ F_0$ where $F_t$ denotes the cumulative distribution function (CDF) of $Y_{C,t}$ for $t=0,1$. Estimation of $\sd$ from data and plugging that estimator into Equation \ref{eq:counterfactual_pushforward} to yield an estimated counterfactual measure is the key idea of the CiC estimator. As previously noted, this estimator can be extended to multivariate outcomes through tensorization, which applies it independently to each coordinate. Joint structure between the coordinates is incompatible with the scalar distribution and quantile functions used in the estimator, as well as with the assumption of scalar production functions.

Recently in the univariate setting, other approaches for modeling the control group time trend have been proposed which generalize the ``parallel trends'' assumption to allow for general heterogeneity. These models in general make stronger or less interpretable assumptions than the monotone production functions of \citet{athey2006identification}. \citet{callaway2019quantile} directly assume that the copulas between $\Prob_{Y_{C,0}}$ and $\Prob_{Y_{T,0}}$ as well as $\Prob_{Y_{T,0}}$ and $\Prob_{{Y}_{T,1}^\dagger}$ are the same; \citet{roth2020parallel} assume that the pointwise differences between the corresponding cumulative distribution functions are equal, i.e.~that $F_1(x)-F_0(x)={F}_1^\dagger(x)-F_0^\star(x)$ for all $x\in\mathbb{R}$; \citet{bonhomme2011recovering} restrict the heterogeneity of the model to be additively separable and assume that the pointwise differences between the corresponding logarithms of the characteristic functions are equal. In the next section we introduce cyclical comonotonicity, a natural extension of the monotonicity requirement in \citet{athey2006identification} to multivariate settings, and hence the multivariate notion of parallel trends closest to the framework of that work.

In the following section, we show that under the CiC model the univariate natural trend map $\sd$ is unique and therefore identifiable as an optimal transport map between $\Prob_{Y_{C,0}}$ and $\Prob_{Y_{C,1}}$. We leverage this insight to identify unique natural trend maps in higher dimensions under the same model. In particular, we propose an assumption of cyclically comonotonicity, a multivariate notion of monotonicity between the production functions, which extends the CiC estimator's assumption of monotone increasing production functions.

\section{Higher Dimensional Changes-in-Changes}

\subsection{An optimal transport extension of the changes-in-changes estimator}\label{sec:CIC_OT}

Identifiability of the causal model presented in the previous section hinges on the uniqueness of the monotone increasing map $\sd$ such that $\sd_{\#} \Prob_{Y_{C,0}} = \Prob_{Y_{C,1}}$. This uniqueness follows from a fundamental result in the theory of optimal transportation known as Brenier's theorem.

\begin{theorem}[\citeauthor{brenier1991polar}, \citeyear{brenier1991polar}]
\label{thm:brenier}
Let $P$ and $Q$ be two probability measures defined over $\R^d$ such that $P$ is absolutely continuous with respect to the Lebesgue measure. Then, among all maps $T: \R^d \to \R^d$ such that $Q=T_{\#}P$, there is a unique one, called the \emph{Brenier map} denoted $\bar T$, which is the gradient of a convex function. Furthermore $\bar T$ is an optimal transport map in the following sense.

Let $\Gamma$ denote the set of joint probability distributions of $(X,Y) \in \R^d\times \R^d$ such that $X\sim P$ and $Y\sim Q$, then the optimal transport problem
\begin{equation}
    \label{eq:OT}
    \inf_{\gamma \in \Gamma} \int \|x-y\|^2 \ud \gamma (x,y)\,,
\end{equation}
admits a unique solution $\bar \gamma$ such that $(X,Y) \sim \bar \gamma$ if and only if $X \sim P$ and $Y= \bar T(X)$, $P$-almost surely. 
\end{theorem}
For $d=1$, a map $T$ is the gradient of a convex function if and only if it is non-decreasing. The quantile map $F_{Y_{C,1}}^{-1}\circ F_{Y_{C,0}}$ has the appropriate domain and range measures and is non-decreasing by the properties of CDFs; therefore it is the unique optimal transport solution.

Theorem \ref{thm:brenier} demonstrates that the structural properties of the univariate CiC model can be extended to higher dimensions by leveraging the theory of optimal transportation. In particular, the Brenier map is unique and hence identifiable between pairs of sufficiently regular measures in $\mathbb{R}^d$. We therefore introduce regularity assumptions on the production functions $h_{\cdot,\cdot}$ in the model which imply that the optimal drift is the Brenier map between $\Prob_{Y_{C,0}}$ and $\Prob_{Y_{C,1}}$. The counterfactual distribution $\Prob_{Y_{T,1}^\dagger}$ in Figure \ref{fig:connectionpic} can be identified by first estimating the unique natural trend $\sd$ between the observable measures $\Prob_{Y_{C,0}}$ and $\Prob_{Y_{C,1}}$ in the control group and then applying that transformation to the treatment group pre-intervention potential outcome measure $\Prob_{Y_{T,0}}.$ 

For our main result, we assume that all four the potential outcome measures $\Prob_{Y_{\cdot,\cdot}}$ and the counterfactual measure $\Prob_{Y_{T,1}^\dagger}$ are absolutely continuous with respect to the Lebesgue measure. This assumption for pre-intervention measures is needed to apply Theorem \ref{thm:brenier}, and for the post-intervention measures it helps enable identification. To satisfy technical assumptions on the regularity of the optimal transport problem that often arise in classical results ~\citep[see, e.g.,][]{caffarelli2000contraction}, we enforce through Assumption \ref{ass:measures} that measures are supported on convex sets. Furthermore, if the support $K_0^\star$ of $\Prob_{Y_{T,0}}$ is not contained in the support $K_0$ of $\Prob_{Y_{C,0}}$, one cannot infer the counterfactual distribution $\Prob_{Y_{T,1}^\dagger}$ at the parts of $K_0^\star$ that fall outside $K_0$ without extrapolation via structural assumptions. This restriction on the supports is a standard assumption that prevents the need for extrapolation \citep{athey2006identification}. We summarize our structural assumptions as follows:

\begin{assumption}\label{ass:measures}
The observable measures $\Prob_{Y_{C,t}}$, $\Prob_{Y_{T,t}}$, $t=0,1$, and the counterfactual measure $\Prob_{Y_{T,1}^\dagger}$ are supported on proper convex subsets $K_t$, $K_t^\star$, and $K^\dagger_1$ of $\mathbb{R}^d$ and are absolutely continuous with respect to Lebesgue measure. Moreover, $K_0^\star\subset K_0$.
\end{assumption}

When $\sd$ is an optimal transport map, Theorem \ref{thm:brenier} implies $\sd$ is the gradient of a convex function. Recall that in higher dimensions, gradients of convex functions from $\mathbb{R}^d\to\mathbb{R}$ are \emph{cyclically monotone} \citep[Theorem 24.8]{rockafellar1997convex}. A map  $T: \mathbb{R}^d\to\mathbb{R}^d$ is cyclically monotone if for any positive integer $m$ and any cycle $u_1, \ldots, u_m, u_{m+1} \equiv u_1$ in its domain, it holds

\begin{equation}
    \label{eq:CM}
    \sum_{i=1}^m\langle u_{i}, T(u_{i})- T(u_{i+1})\rangle \ge 0. 
\end{equation}

Therefore, the natural trend map is identifiable when it is cyclically monotone. For $m=2$, Equation~\eqref{eq:CM} reduces to the simple notion of multivariate monotonicity defined as
\begin{equation}
\label{eq:monotone}
  \langle u_1 -u_2, T(u_1)-T(u_2)\rangle \ge 0\,.  
\end{equation}
Furthermore, in the univariate case these two definitions are equivalent; in the CiC model this implies that the monotone map $h_1 \circ h_0^{-1}$ is also cyclically monotone, which explains why the assumption of monotonicity of the functions $h$ in the unobservable is sufficient.

In the context of our latent model, the monotonicity assumption on the production functions for the CiC estimator implies that larger values of the unobservable latent variable correspond to strictly larger potential outcomes. This is a common assumption in economic models \citep{matzkin2003nonparametric, chernozhukov2005iv, imbens2007nonadditive}, but can be restrictive in some settings. For instance, consider a classroom experiment where the treatment variable is class size and the outcomes of interest are standardized test scores before the intervention and 5 years after. Monotone production functions imply that individuals with higher unobserved ability $U$ will receive strictly higher test scores. Specifically, the assumption of being able to rank individuals in terms of one univariate unobservable $U$ is a strong restriction.

A careful inspection of the data generating process reveals that monotone assumptions on the production functions allow the latent variable to be entirely abstracted away. This actually follows from the weaker notion of comonotonicity such that $h_0,h_1:\R^d \to \R^d$ are comonotone if $\langle h_0(x)-h_0(y),h_1(x)-h_1(y)\rangle\geq 0$ for all $x,y \in \R^d$. Note that with an identity production function comonotonicity reduces to classical monotonicity.
To get a sense of the strength of this relaxation, consider univariate differentiable functions. In this case comonotonicity implies locally that their derivatives $h_0'\cdot h_1'\ge 0$ have the same sign and furthermore imposes additional global constraints. As a concrete example, if $h_0$ is a polynomial production function and $h_1$ its pointwise scaling by some $\gamma > 0$, then this pair of functions is comonotone because they have the same sign between all zeros and hence the same signed difference between any pair of points. This example emphasizes that $h_0$ and $h_1$ need not be individually monotone themselves.

We have now seen that that the monotone production functions assumption in \citet{athey2006identification} implies that the natural trend is cyclically monotone, which enables identification of the counterfactual. Furthermore, the production functions are comonotone, which allows the model to include general latent space distributions. However, these two functional assumptions only collapse to monotonicity in the scalar case. To extend their model to higher dimensions, we propose an assumption of cyclically comonotone production functions which achieves both these desired properties. \begin{definition}Two production functions $h_0$ and $h_1$ are \emph{cyclically comonotone} if for any positive integer $m$ and any cycle $u_1, \ldots, u_m, u_{m+1}=u_1$ in their common domain, it holds
\begin{equation}
    \label{eq:CCM}
    \sum_{i=1}^m\langle h_0(u_{i}), h_1(u_{i})- h_1(u_{i+1})\rangle \ge 0. 
\end{equation}
\end{definition}
Just as cyclical monotonicity collapses to monotonicity when $m=2$, cyclical comonotonicity collapses to comonotonicity in that case. Whenever $h_0$ has an inverse $h_0^{-1}$, condition~\eqref{eq:CCM} implies the map $\sd=h_1 \circ h_0^{-1}$ is also cyclically monotone. This result is shown in the proof of Theorem \ref{thm:main}, which we will state shortly. By Theorem~\ref{thm:brenier}, $\sd$ is the unique Brenier map such that $\Prob_{Y_{C,1}}=\sd_{\#}\Prob_{Y_{C,0}}$.

Cyclical comonotonicity is one of several concepts to extend comonotonicity to higher dimensions. One set of extensions imposes a total ordering on the outcome space; however, these imply that the copula of outcomes pre- and post-intervention are identical, a strong assumption on the causal model. Note that the tensorized CiC assumes this shared copula structure. More recently, optimal transportation theory has been used to propose other multivariate notions of comonotonicity. A particular contribution is $\mu$-comonotonicity from \citet{galichon2010comonotonic}. Similar to our model, the authors assume a latent variable and production functions mapping it to potential outcomes. Both production functions $h_0$ and $h_1$ are assumed to have optimal transport structure, as they lie in the graph of a gradient of a convex function. In general, the composition of optimal transport maps, such as $h_1 \circ h_0^{-1}$, is not an optimal transport map, except in special circumstances \citep{boissard2015distribution}. \citet{puccetti2010multi} provide an overview of these ideas.

Cyclical monotonicity is a natural extension of monotonicity to multivariate settings with many uses in economic theory and mechanism design \citep{ashlagi2010monotonicity, rochet1987necessary}, econometrics \citep{shi2018estimating, chernozhukov2021identification}, and statistics \citep{ruschendorf1996c}. Cyclical comonotonicity is even weaker than this, as it does not imply that each production function itself is cyclically monotone, which makes it a natural and weak extension of the monotonicity assumption in the univariate setting. In particular, it is weaker than the assumption that individuals can be ranked in terms of a univariate unobservable variable (ability in the classroom example) and that this ranking persists in the observed outcome distribution. We argue that this makes it a more natural assumption than the monotonicity assumption of the CiC estimator in practical settings.
Examples of cyclic monotonicity in causal inference have been highlighted in \citet{chernozhukov2021identification}, \citet{shi2018estimating}, and \citet{gunsilius2022condition}, among many others. More fundamentally, \citet{rochet1987necessary} shows that the maximizer of quasi-linear utility functions is cyclically monotone. For a concise survey, we refer to \citet{galichon2017survey}.

From a practical perspective, cyclical comonotonicity is an important weakening of cyclic monotonicity in that it allows for unobservables of general dimension in practice. Allowing for multivariate unobserved heterogeneity in general causal models has been a long-standing challenge: in most models, the unobservable $U_{i,t}$ is required to be univariate \citep{athey2006identification, matzkin2003nonparametric, imbens2007nonadditive}. By relying on the concept of cyclical comonotonicity, we can relax this assumption by only requiring that the unobservable is of the same dimension as the observable variable. Note that even this can be relaxed by using recent advances in optimal transport theory \citep{mccann2020optimal}. The cost of this relaxation is a more complicated model and definitions from geometric measure theory like the area and coarea formula, which is why we focus on the simpler setting of unobservables of the same dimension in this article.

\paragraph*{Discrete outcome distributions}
Mirroring the univariate setting \citep{athey2006identification}, our methodology does not guarantee a point identified-counterfactual distribution when outcomes are discrete-valued. Theorem \ref{thm:brenier} guarantees a unique control group natural trend with optimal transport structure when the measure of pre-intervention outcomes is absolutely continuous with respect to the Lebesgue measure. Indeed, unique optimal transport maps between discrete measures need not exist. For illustration, consider transporting a measure distributed on vertices of the unit square with equiprobability at each main diagonal vertex to a one with equiprobability at the anti-diagonal vertices. With respect to squared distance, it is equivalent to transport the mass between vertical or horizontal pairs of vertices. Furthermore, it is admissible to split mass and transport $0.25$ probability mass from each main diagonal vertex to each anti-diagonal vertex. This solution with fractional structure is an example of a non-deterministic optimal transport \textit{plan} (cf. deterministic optimal transport \textit{maps}), which can be interpreted as a probabilistic assignment rule. Discrete-valued optimal transport problems often have a minimum cost achieved by multiple probabilistic plans but no deterministic maps. This reality implies that point-identification in the discrete case need not be possible in general without restrictive further assumptions.

This optimal transport view hence also explains the lack of point-identification in the univariate setting, matching the conclusions of \citet{athey2006identification}. The extension of CiC to discrete outcomes in that setting provides bounds on the counterfactual outcome distribution due to the same challenges described above. Their estimator for that case is heavily dependent on notions of quantile functions without a direct multivariate extension. Identification only becomes possible with additional assumptions about conditional independence and exogenous covariates, which the authors admit are restrictive. In the discrete multivariate setting, similar partial identification results are possible through carefully application of recent developments in optimal transport theory. For example \citet{auricchio2022structure} bound the supremum cost of moving any single point mass in $\ell^p$-cost transport problems. With one possible counterfactual distribution identified by a solution to the minimization, uniform bounds can then be found by applying the aforementioned bound on the possible pointwise deviation. Point-identification can also be reestablished in a similar manner to the univariate setting by introducing additional information from covariates under strong assumptions. The focus of this article is to introduce an optimal transport framework for point-identification, which is why we focus on the case of absolutely continuous measures throughout.

\subsection{Identification of causal effects in the multivariate setting}\label{sec:identification}
Based on the extensions of the causal model introduced in the previous sections, we can formally state the identification result for the counterfactual distribution of the treated units had they not received treatment. 

\begin{theorem}
\label{thm:main}
Consider the causal model depicted in Figure \ref{fig:connectionpic}. Let Assumption \ref{ass:measures} hold. Moreover, assume that the production function $h_0$ has a well-defined inverse and that $h_0$ and $h_1$ are cyclically comonotone in the sense of~\eqref{eq:CCM}. Then there exists a unique map $\sd: K_0^\star \to K_1^\dagger$. It is the Brenier map from $\Prob_{Y_{C,0}}$ to $\Prob_{Y_{C,1}}$. The counterfactual distribution $\Prob_{Y_{T,1}^\dagger}$ of the treated unit had it not received treatment is then identified via $\Prob_{Y_{T,1}^\dagger} = \sd_{\#}\Prob_{Y_{T,0}}$.
\end{theorem}
The proof of Theorem \ref{thm:main} is relegated to Appendix \ref{sec:proof_main}. 
Another potential quantity of interest is the actual counterfactual random variable $Y_{T,1}^\dagger$. Identifying this quantity requires more structure in the model. Indeed, in the setup of the causal model represented in Figure \ref{fig:connectionpic}, we only have that $\Prob_{Y_{C,1}}=\sd_\# \Prob_{Y_{C,0}}$ but not necessarily that $Y_{C,1}=\sd (Y_{C,0})$ because we do not assume\footnote{We note in passing that if we were to assume that $U_{i,0}=U_{i,1}$ for all units, we could, in fact, test the assumption that the natural trend $\sd$, and hence the production functions $h_0,h_1$, are monotone increasing from independent copies of paired data of the form $(X_0,Y_1)$. We would do so by checking if there exists an increasing function $\sd$ that indeed interpolates these points. In practice, such a function seldom exists and allowing a time-variable $U_{i,t}$ accounts for deviations from this ideal situation.} that $U_{i,0}=U_{i,1}$ for all units. However, assuming that both $h_1^\dagger$ and $h_1$ are cyclically comonotone allows to identify $Y_{T,1}^\dagger$, which follows from the fact that $T:K_1^\star\to K_1^\dagger$ will be identified for the fixed unobservable $U_{i,1}$ for treated units.

\begin{corollary}\label{corr:main}
Consider the setting and assumptions from Theorem \ref{thm:main}. If the production function $h_1$ has a well-defined inverse and $h_1^\star$ and $h_1$ are cyclically comonotone, then there exists a unique map $\sT: K_1^\star \to K_1^\dagger$ such that $\Prob_{Y_{T,1}^\dagger} = \sT_{\#} \Prob_{Y_{T,1}}$. $\sT$ is the Brenier map from $\Prob_{Y_{T,1}}$ to $\Prob_{Y_{T,1}^\dagger}$. $Y_{T,1}^\dagger$ is then identified via $Y_{T,1}^\dagger = T(Y_{T,1})$.
\end{corollary}
The assumption that $h_1$ and $h_1^\star$ are cyclically comonotone can be hard to satisfy in practice, which is why the main parameter of interest is the counterfactual law $\Prob_{Y_{T,1}^\dagger}$ and not the random variable $Y_{T,1}^\dagger$. It is worth noting that having cyclically comonotone pairs $h_1^\star$ and $h_1$ as well as $h_0$ and $h_1$ does not necessarily imply $h_1^\star$ and $h_0$ are cyclically comonotone.

In general, the assumptions of our model match those in \citet{athey2006identification} when outcomes are scalar and extend their structure in higher dimensional cases. Our production function setup illustrated in Figure \ref{fig:connectionpic} exactly matches that of CiC. In the scalar case, our assumption of comonotone production functions reduces to monotone production functions; although this is weaker than CiC's requirement that $h_0$ and $h_1$ are strictly monotone, the requirement of Theorem \ref{thm:main} that $h_0$ has a well-defined inverse restricts that pre-intervention production function to be strictly monotone as well. In one dimension, a strictly monotone and a weakly monotone production function pair remains cyclically comonotone, so our model does provides a relaxation of the control group outcome generation process. Here it is the assumption that $h_0$ has an inverse which forces the production functions to be themselves monotone, as contrasted by our previous example of two positively proportional polynomials being comonotone. In higher dimensions, however, the invertibility of $h_0$ does not imply both $h_0$ and $h_1$ are themselves cyclically monotone. If this were the case, then both production functions would be optimal transport maps, which can be seen by considering cyclical monotonicity as the special case of cyclical comonotonicity when $h_0$ is the identity. This is precisely the notion $\mu$-comonotonicity introduced by \citet{galichon2010comonotonic}.

For the latent distributions, our extension retains the time-invariance within groups assumption of CiC.  While we do not make the same assumption of nested latent supports $\nu^\star \subset \nu$ as \citet{athey2006identification}, our assumption that $K_0^\star \subset K_0$ is equivalent. To see why note that both treatment arm share the same pre-intervention production function $h_0$. Our assumption that the support sets are convex is an additional requirement not needed for CiC but necessary to apply theoretical results about multivariate optimal transport. \citet{athey2006identification} allow for an additional case with discrete latent distributions, but our assumption that $h_0$ is invertible with a continuous co-domain implies that the latent variable measures must be continuous.

\paragraph*{Examples of functionals that can be analyzed with the proposed method}
One of the advantages of the method is to deal with multivariate outcome distributions.
Many functionals of interest in different areas of the (social) sciences are based on the information of the joint distribution of several outcomes of interest, from social inequality \citep{decancq2013weights, duclos2006robust, bourguignon2003measurement} to risk management \citep{ekeland2012comonotonic, embrechts2002correlation} to decision theory \citep{hallin2022measure}, to name a few.

More generally, the proposed estimator provides consistent estimates of functionals that fundamentally rely on the \emph{copula structure} of the outcome distributions. Examples of this are multivariate stochastic dominance \citep{garcia2019review}, multivariate risk \citep{embrechts2006bounds}, and functionals that encompass interrelation between random variables such as mutual information or correlation. 
We say that a distribution $F$ dominates another distribution $G$ in the $\alpha$-th order, written as $F\succsim G$ if $D_\alpha^F(x)\leq D_\alpha^G(x)$, where
\[D_1^F(x)\equiv F(x) \qquad\text{and}\qquad D_\alpha =\int_{S(x)} D_{\alpha-1}(t)d t,\] with $S(x)\coloneqq \left\{y\in \mathbb{R}^d_+: y\leq x\right\}$, see \citet{garcia2019review} for details. An example of a multivariate risk measure is the multivariate lower-orthant value-at-risk at level $\alpha\in[0,1]$ of a multivariate increasing function $G$, which is defined as \citep{embrechts2006bounds} 
\[\underline{\text{VaR}}_\alpha(G) = \partial\left\{x\in\mathbb{R}^d: G(x)\geq\alpha\right\},\] where $\partial A$ denotes the boundary of the set $A$. An analogous definition holds for the upper-orthant counterpart. The mutual  information $I(X,Y)$ of two random variables $X$ and $Y$ is defined as \citep{shannon1948mathematical}
\[I(X,Y) = \text{KL}\left(P_{X,Y} || P_X\otimes P_Y\right),\] where $\text{KL}$ is the Kullback-Leibler divergence and $P_X\otimes P_Y$ is the independence coupling.

The above examples are only a short list of potentially interesting functionals one can analyze using the proposed method. All of these functionals rely on the information of the corresponding copula structure, which cannot be identified with a tensorized version of the changes-in-changes estimator unless in the limited setting of completely independent coordinates described previously.

\subsection{Estimation from Observed Data}\label{sec:estimation}
In practice we only observe observations from each of the four distributions $\Prob_{Y_{C,0}}, \Prob_{Y_{C,1}}, \Prob_{Y_{T,0}}$, and $\Prob_{Y_{T,1}}$. In order to distinguish between population measures and empirical measures, we define the corresponding empirical measures of $\Prob_{Y_{C,0}}$ as $\hat{\mu}_{0,n}$, of  $\Prob_{Y_{C,1}}$ as $\hat{\mu}_{1,l}$, and of $\Prob_{Y_{T,0}}$ as $\hat{\mu}_{0,m}^\star$ with corresponding sample sizes $n,l,m$. We denote the estimator of the counterfactual $\Prob_{Y_{T,1}^\dagger}$ by $\hat{\mu}_{1,m}^\dagger$. Estimating Brenier maps from data is a subject of intense activity both from theoretical and practical angles~\citep{forrow2019statistical, gunsilius2021convergence, hr2021minimax, delara2021consistent, ManBalNil21,manole2021sharp,pooladian2021entropic,VacMuzRud21,muzellec2021near,deb2021rates}. Further details about estimating Brenier maps from data can be found in Appendix \ref{sec:estimation_from_data}. In this section we state propositions about the consistency of the estimated counterfactual distribution and its rate of convergence in expected Wasserstein distance to the population distribution. The $2$-Wasserstein distance between measures $\mu$ and $\nu$, which we denote as $W_2\left(\mu,\nu\right)$, is the square root of the infimum transportation cost \eqref{eq:OT} between the two measures. Proofs of these propositions are deferred to Appendix $\ref{sec:proofs}$. 

The observations form empirical measures $\hat{\mu}_{0,n}$, $\hat{\mu}_{0,m}^\star$, and $\hat{\mu}_{1,l}$ with $n$, $m$, and $l$ denoting the respective sample sizes. These samples are assumed to be mutually independent. An estimator for the natural trend, $\sdhat$ is a minimizer of $W_2(\hat{\mu}_{0,n}, \hat{\mu}_{1,l})$. A plug-in estimator for the counterfactual distribution is given by $\sdhat_{\# \hat{\mu}_{0,m}^\star}$. However, $\sdhat$ is non-trivially defined only on the support of  $\hat{\mu}_{0,n}$. Therefore in practice we project $\hat{\mu}_{0,m}^\star$ onto the support of $\hat{\mu}_{0,n}$ through one-nearest neighbor Euclidean matching, itself a 2-Wasserstein optimal transport problem. The ``rounded'' measure of treatment outcomes pre-intervention is denoted $\tilde{\mu}_{0,m}^\star$. This leads to our proposed estimator for the counterfactual distribution: ${\sdhat_{\# \tilde{\mu}_{0,m}^\star} =: \hat{\mu}_{1,m}^\dagger}$.

Under the assumptions on population measures stated in Assumption \ref{ass:measures}, the Wasserstein distance between the estimated and population counterfactual distribution converges uniformly to zero. This enables a consistency statement about that distance converging to probability to zero, as well as a stronger statement about the Wasserstein-distance of the estimated and population natural trend maps.

\begin{proposition}\label{prop:consistency}
    Under the conditions of Assumption \ref{ass:measures}, 
    \[\limsup_{\min(m,n) \to \infty} \int_{K_0^\star} \lVert \sdhat(\tilde{x}) - \sd(x) \rVert \Prob_{Y_{T,0}}(dx) =0.\] Furthermore, the squared 2-Wasserstein distance of the empirical natural trend map, $W_2^2(\hat{\mu}_{0,m}^\star, \hat{\mu}_{1,m}^\dagger)$, converges in probability to the distance from pre-intervention samples to their population counterfactual, $W_2^2(\hat{\mu}_{0,m}^\star, \Prob_{Y_{T,1}^\dagger})$. 
\end{proposition}

Under mild further assumptions, we show that the expected 2-Wasserstein distance between the population and estimated counterfactual measures converges to $0$ at a $O(\text{min}(m,n)^{\nicefrac{1}{d}})$ rate.

\begin{assumption}\label{ass:rate}
    The convex support $K_0$ is compact and additionally the Radon-Nikodym derivative $f\coloneqq \frac{d\Prob_{Y_{T,0}}}{d\Prob_{Y_{C,0}}}$ of $\mu_0^\star$ with respect to $\mu_0$ is bounded.
\end{assumption}

Combining these assumptions with those previously stated for consistency under Assumption \ref{ass:measures} yields the following claim.

\begin{proposition}\label{prop:rate}
    The expected 2-Wasserstein distance between $\hat{\mu}_{1,m}^\dagger$ and $\Prob_{Y_{T,1}^\dagger}$ converges to $0$ at a $O(\text{min}(m,n)^{\nicefrac{1}{d}})$, that is 
    \[\E_{\Prob_{Y_{C,0}}, \Prob_{Y_{T,0}}, \Prob_{Y_{C,1}}} W_2(\hat{\mu}_{1,m}^\dagger, \Prob_{Y_{T,1}^\dagger}) = O(\text{min}(m,n)^{\nicefrac{1}{d}}).\]
\end{proposition}

Obtaining the large sample distribution of optimal transport maps is largely an open problem. However, rates of convergence are known \citep[e.g.,][]{deb2021rates, hr2021minimax}, which enables inference on these functionals through subsampling methods. Under mild assumptions \citep[e.g., those in ][]{politis2001asymptotic, romano2012uniform, shao2013fixed}, subsampling theory provides confidence regions and hypothesis tests with strong asymptotic guarantees. Even if the rate of convergence for the functional is unknown, that rate may still be estimated.

\section{Numerical Experiment and Application to the Card \& Krueger Dataset}\label{sec:application}
In this section, we demonstrate the performance of the optimal transport-based estimator with two numerical experiments and apply it to a classical causal inference dataset.

\subsection{A stylized example}\label{sec:motivating_example}
We first present a simple simulation experiment which demonstrates that improper modeling of dependency between coordinates of multivariate outcomes can lead to highly biased estimates. In particular, our proposed method manages to estimate the correct bivariate counterfactual distributions, while the multivariate extension of the CiC estimator through tensorization does not. Consider the following set of linear production functions in which map latent vectors in $\mathbb{R}^2$ to outcomes without intervention in $\mathbb{R}^2$:
\[h_0(u) = \begin{bmatrix} 1&\alpha\\\alpha& 1\end{bmatrix}u\qquad\text{and}\qquad h_1(u) = \begin{bmatrix} 1&-\alpha\\-\alpha& 1\end{bmatrix}u.\]

\begin{figure}[ht!]
    \centering
     \begin{minipage}[b]{0.45\textwidth}
         \centering
         \includegraphics[width=\textwidth]{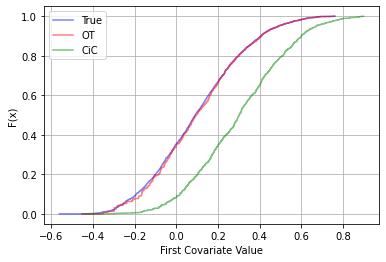}\\\vspace{.2cm}
         (a) First Marginal
         \label{fig:first_marginal}
     \end{minipage}
     \begin{minipage}[b]{0.45\textwidth}
         \centering
         \includegraphics[width=\textwidth]{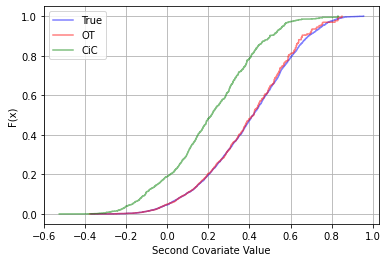}\\\vspace{.2cm}
         (b) Second Marginal
         \label{fig:second_marginal}
     \end{minipage}
     \hfill
        \caption{Recovery of counterfactual marginals by OT and CiC.}
        \label{fig:marginal_recovery_plots}
\end{figure}    

\begin{figure}[ht!]
    \centering
     \begin{minipage}[b]{0.3\textwidth}
         \centering
         \includegraphics[width=\textwidth]{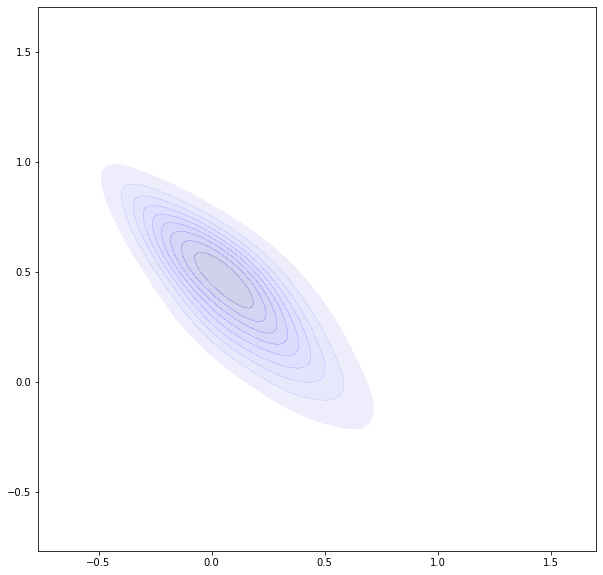}\\\vspace{.2cm}
         (a) {True CF}
         \label{fig:kde_true}
     \end{minipage}
     \begin{minipage}[b]{0.3\textwidth}
         \centering
         \includegraphics[width=\textwidth]{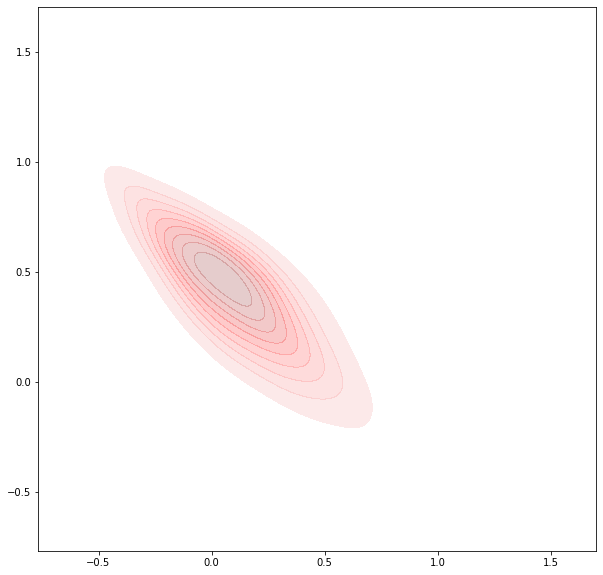}\\\vspace{.2cm}
         (b) OT Estimate
         \label{fig:kde_ot}
     \end{minipage}
     \begin{minipage}[b]{0.3\textwidth}
         \centering
         \includegraphics[width=\textwidth]{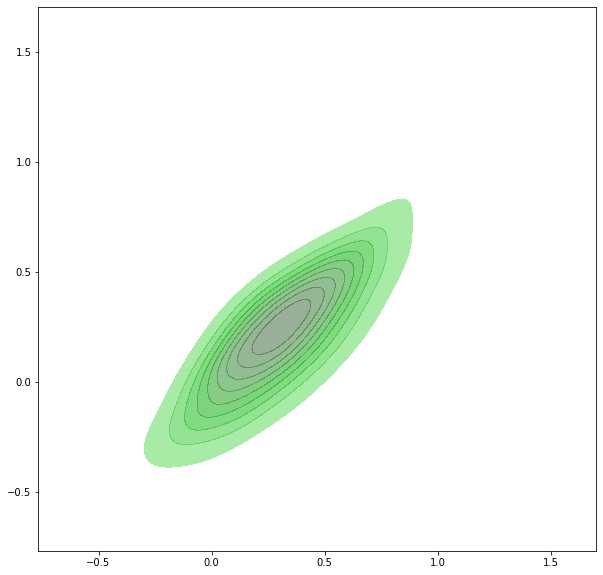}\\\vspace{.2cm}
         (c) CiC Estimate
         \label{fig:Estimate}
     \end{minipage}
     \hfill
        \caption{KDE plots of the distribution of counterfactual outcomes for the treatment group. The first covariate increases along the horizontal axis and the second along the vertical. KDE bandwidth = 0.5.}
        \label{fig:kde_plots}
\end{figure}    

These production functions are a cyclically comonotone but not element-wise monotone as required for the CiC estimator. Choosing Beta distributions as the laws of the latent variables, we generate $n=3000$ independent outcomes from the control population before and after the intervention, as well as from the treatment group before the intervention and its unobserved counterfactual. Further details about the implementation can be found in Appendix \ref{sec:motivating_example_details}.

The recovered marginals in Figure \ref{fig:marginal_recovery_plots} and the kernel density estimator (KDE) of the counterfactual joint distribution in Figure \ref{fig:kde_plots} demonstrate that our methodology estimates the true counterfactual distribution almost exactly while the CiC estimator cannot. Notably, the CiC estimator recovers a joint distribution with a mirrored dependence structure.

As previously noted, the tensorized CiC estimates counterfactuals under the strong assumption that the copula of outcomes pre- and post-intervention are identical. Each coordinate of the potential outcomes vectors is assumed to be a monotone increasing function of the latent random vector's corresponding coordinate. When there are interactions between latent coordinates during potential outcome generation the multivariate CiC will not identify counterfactual distributions. We believe that proper accounting for correlation structures contributes to our seemingly novel substitution result when reanalyzing the classical minimum wage data from \citet{card1994minimum} in Section \ref{sec:ck_reanalysis}.
 
\subsection{Numerical Experiment in Higher Dimensions}\label{sec:numerical_experiment}
\begin{figure}[h]
    \centering
    \begin{minipage}[b]{0.5\textwidth}
    \centering
    \includegraphics[width=\textwidth]{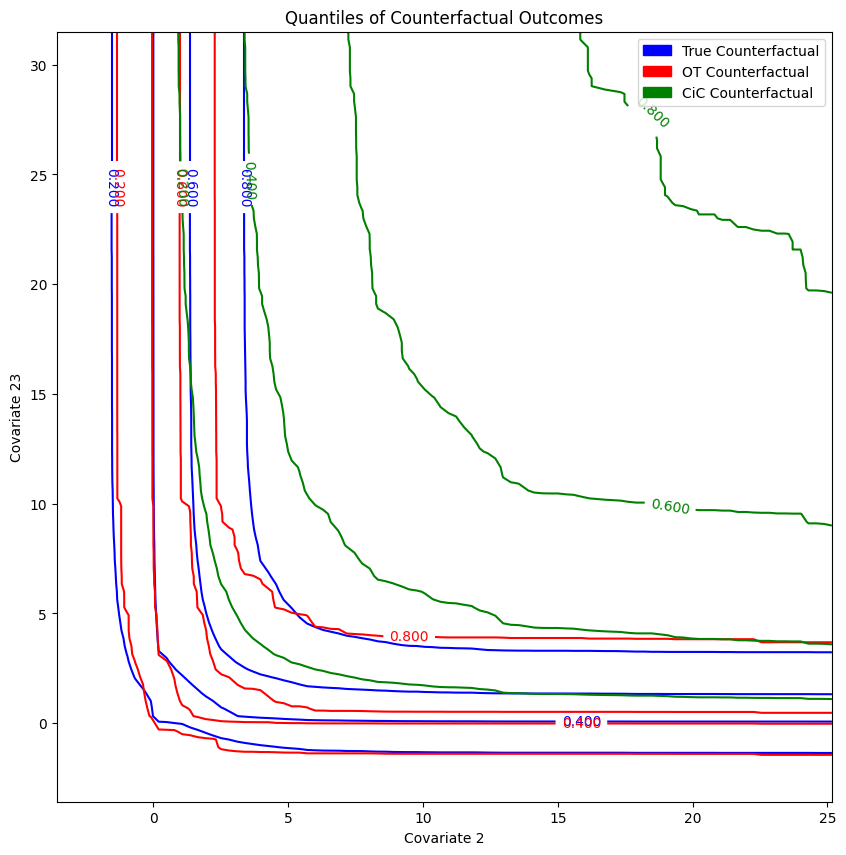}
    \caption{True and estimated empirical CDF quantiles for counterfactual treatment individuals, D=50}
    \label{fig:numerical_experiment_joint}
    \end{minipage}
\end{figure}

In the following numerical experiment, we consider the performance of our estimator against the multivariate tensorized CiC estimator as the dimensionality of the observations grows. Optimal transport's strong recovery of higher-dimensional joint distributions, illustrated by Figure \ref{fig:numerical_experiment_joint}, distinguishes it.

Recall from Theorem $\ref{thm:brenier}$ that optimal transport maps can be characterized as the gradients of convex functions. Consider the function $f(x,y) = x^2 y^{-1}$ which is convex when $y > 0$. As the sum of convex functions is convex, summing $f$ evaluated at multiple pairs of coordinates in a random vector constructs a convex function from $\mathbb{R}^d$ to $\mathbb{R}$. Formally we consider a class convex functions $g(x; \theta) = \sum_{i=1}^p x_{(i_1)}^2  x_{(i_2)}^{-1}$ where $x \in \mathbb{R}_{++}^d$, $x_{(i)}$ represents the $i^{\text{th}}$ coordinate of $x$, and $i_j \in [[d]]$ for $j \in \{1, 2\}$; the parameter $\theta$ contains the $p$ pairs of indices to be compared. The gradient of $g$ is straightforward to compute. In this simulation, we specify $h_0$ to be the identity and $h_1$ to be $\nabla g$. There are $d$ coordinate pairs randomly selected each run. Both treatment arms have $n=3,000$ units drawn from Beta latent distributions. Further implementation details can be found in Appendix \ref{sec:numerical_experiment_details}.

Our experiment is designed quantify how well the two methods estimate a joint counterfactual distribution. For computational reasons, we focus on randomly selected pair-wise coordinate interactions. To this end, we construct the empirical CDF (eCDF) of the true counterfactual, its optimal transport estimate, and its CiC estimate over a uniform mesh of $10,000$ points. We compute the mean absolute difference between the true and estimated eCDFs over each point in the mesh and have $20$ runs per trial. As demonstrated in Figure \ref{fig:numerical_experiment_joint}, our proposed method better estimates the quantiles of a plane in the outcome space. Table \ref{table:experiment_results} shows that optimal transport has a mean absolute difference consistently smaller than changes-in-changes with a smaller standard deviation as well. Our method's monotone increasing loss likely occurs due to the curse of dimensionality and the difficulty of matching vectors in very high dimensions with only $3000$ samples. Regardless, optimal transport still has substantially smaller MAD and standard deviation than CiC when $d=100$.
\begin{table}
    \centering
    \caption{Mean absolute deviation of the estimated counterfactual CDF to the true one. eCDFs are evaluated over one randomly selected plane per run.}
    \label{table:experiment_results}
    \begin{tabular}{c|c|c|c|c|c}
    Dimension &2&10&20&50&100\\\hline\hline
    OT MAD&$.0006$&$.0017$&$.0019$&$.0033$&$.0037$\\
    st.~dev. &$.0003$&$.0011$&$.0016$&$.0027$&$.0055$\\\hline
    CiC MAD&$.0083$&$.0104$&$.0092$&$.0088$&$.0090$\\
    st.~dev.&$.0055$&$.0078$&$.0074$&$.0073$&$.0125$\\\hline
    \end{tabular}
\end{table}

\subsection{Revisiting the Card \& Krueger dataset}\label{sec:ck_reanalysis}
On April 1, 1992, New Jersey raised its minimum wage from the Federal level of $\$ 4.25$ per hour to $ \$ 5.05$ per hour. \citet{card1994minimum} (CK henceforth) investigated the effect this increase had on employment in fast food restaurants with a case-control experimental design, taking the bordering region of eastern Pennsylvania, where the minimum wage remained at $\$ 4.25$, as a control group. The original study employs the standard DiD estimator to conclude that the higher minimum wage led to increased employment in New Jersey restaurants. This result spawned much debate within the economics community about the effect of minimum wage policies on employment (e.g.,~\citet[][\citeyear{neumark2000minimum}]{neumark2000minimum}, \citet[][\citeyear{dube2010minimum}]{dube2010minimum}, \citet[][\citeyear{meer2016effects}]{meer2016effects}, \citet[][\citeyear{neumark2014revisiting}]{neumark2014revisiting}, and references therein).

The treatment effect of interest in CK and many subsequent reevaluations is the change in full-time equivalent employees (FTE), which is defined via ${\text{FTE} = \text{FT} + .5 \text{PT}}$, where FT (PT) is the number of full-time (part-time) employees \citep{neumark2000minimum,lu2004optimal}. We use the proposed multivariate extension of the changes-in-changes estimator to estimate a bivariate treatment effect in terms of full- and part-time employees, hence fully dissecting the causal effect of raising the minimum wage on these subgroups.

We analyze the number of full- and part-time employees as drawn from an underlying continuous distribution. In the original Card and Krueger data, ``there are 28 records that report fractional full-time employees (the fraction is always one-half), 29 records that report fractional numbers of part-time employees, and one record that reports a fractional number of managers'' \citep{ehrenberg1995myth}. These partial observations could represent ambiguity in the classification of a worker as full- or part-time \citep[Alan Krueger via interview reported in][]{ehrenberg1995myth}. As the restaurant employee responding to the phone interview was not necessarily the same at both measurements, \citet{neumark2000minimum} argue, ''there is no reason to believe that the responses in the first and second waves are
based on the same `definition' of employment.'' Therefore these fractional observations can also be considered a degree of belief statement. As discussed above, deterministic optimal transport maps are the unique solution in the continuous case while the discrete case also admits probabilistic optimal transport plans as solutions. In practice we use a linear program implemented in the Python package \texttt{POT} which finds a deterministic plan, matching the structure of our theoretical results.

The original CK study finds a positive average treatment effect of $2.76$ FTE jobs added in the treatment group compared to the control group, a result reproduced by \citet{neumark2000minimum} using different methodology. \citet{neumark2000minimum} also compute an average treatment effect in full- and part-time jobs with separate regressions and find treatment effects of $3.16$ gained full-time jobs and $0.60$ lost part-time jobs. These classical contributions only focus on aggregate outcomes over all restaurants, irrespective of the size of the respective restaurant. An exception is \citet{ropponen2011reconciling} who reanalyzes the data with the CiC estimator, taking into account the heterogeneity in the size of the fast-food restaurants. He finds the average treatment effect bounded in $[0.90,1.70]$. Ropponen also notes in New Jersey that large restaurants pre-intervention (measured in FTE employees) have a negative treatment effect while small restaurants pre-intervention have a positive treatment effect. Our analysis recovers this trend for both FT and PT employees in New Jersey. As we also generate counterfactual controls, we find an opposite trend for Pennsylvania restaurants. This opposite trend is intriguing and worth exploring further as it does not seem to be explained by substitution effects of employees moving across the state line to find jobs.

\begin{table}[h]
\centering
\caption{Estimated Average Treatment Effect on Full- and Part-Time Employment by Method with Subsampled 95\% Confidence Intervals.}\label{table:summarytable}
\scalebox{0.85}{
\begin{tabular}{c|c|c|c}
&OT&CiC&DiD\\\hline\hline
ATE FT &$3.07 \ (2.00, 4.67)$&$2.61 \ (0.40,3.45)$&$3.45\ (3.05,5.46)$\\\hline
ATE PT &$-1.79 \ (-3.43, -0.33)$&$-1.52 \ (-3.46,-0.51)$&$-1.00\ (-2.07,1.15)$\\\hline\hline
\end{tabular}}\vspace{.3cm}
\end{table}

The previously discussed subsampling approach to inference was applied to generate symmetric 95 \% confidence intervals around each point estimate. The only interval to contain zero is for the DiD treatment effect on part-time employment. Furthermore, we run subsampled hypothesis tests to check whether the point estimates are significantly different across estimators. Of the four tests between optimal transport to an alternative, only the difference in treatment effect on part-time employment estimated with CiC was not significantly different from zero at the $p = 0.05$ level. For both the confidence intervals and hypothesis tests, we use a block size of 300 and 10,000 replications.

The average treatment effects estimated using the OT and CiC counterfactual distributions depend only on the marginals, not the full joint structure. However as illustrated in Figure \ref{fig:ck_marginals}, the strong copula assumptions between pre- and post-intervention periods for CiC can lead to markedly different estimated marginals. That structure implies restaurants with the largest pre-intervention full/part-time employment will also have the largest counterfactual full/part-time employment. The wider tails of the CiC treatment effect marginals compared to OT suggest that additional joint structure in this setting leads to counterfactuals with less variance from the observed potential outcome.

The proposed multivariate extension of the changes-in-changes estimator allows us to jointly estimate the effects on full- and part-time employment while accounting for the heterogeneity in restaurant size. For the results in this section, restaurants are represented in $\mathbb{R}^2$ by our outcomes of interest, the number of full- and part-time employees. In the supplementary materials we provide an additional experiment with restaurants represented as higher dimensional vectors including additional measurements about wages, prices, and operations. We discard $19$ units with missing outcomes, leaving $76$ control and $315$ treatment restaurants. The transport plans for both groups are solved with a linear program, and we apply nearest neighbor matching between treatment and control pre-intervention samples to calculate counterfactual outcomes.  

\begin{figure}[h]
    \centering
    \begin{minipage}[b]{0.5\textwidth}
    \centering
    \includegraphics[width=\textwidth]{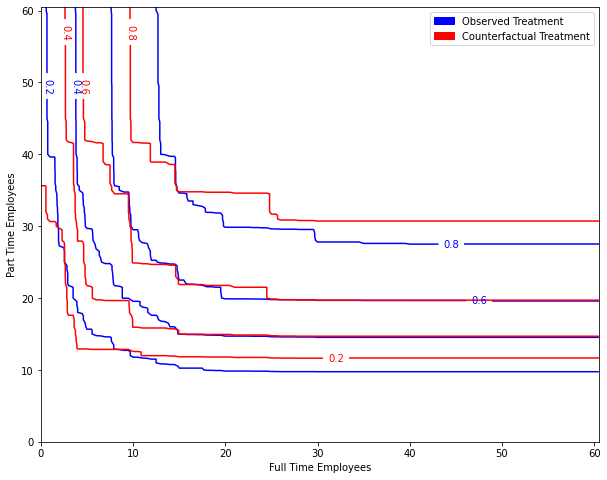}
    \caption{Post-intervention quantile curves for observed treatment group units and their counterfactuals.}
    \label{fig:CK_quantiles}
    \end{minipage}
\end{figure}

Our optimal transport analysis suggests that fast food restaurants responded to an increased minimum wage by substituting full-time employees for part-time ones with a net gain of full-time equivalent employees. Our results seem to indicate that the negative effect on part-time employees is more pronounced than previously estimated. The quantile plot in Figure \ref{fig:CK_quantiles} suggests this effect applies throughout the distribution of restaurants, not just the mean: fixing the number of full-time employees, counterfactual quantiles tend to have more part-time employees than treatment group quantiles at the same level; likewise fixing the number of part-time employees, treatment group quantiles tend to have more full-time employees. In their original paper CK suggest that restaurants may respond to higher labor costs with more full-time employees because they tend to be older, more skilled, and more productive, thus a better investment of capital. Furthermore, the univariate methods may underestimate the number of part-time employees lost, a conclusion supported by the higher dimensional analysis in Table \ref{table:robustness_results} from the supplementary materials. 

\section{Discussion}
We have introduced a general method based on optimal transport theory for causal inference in observational treatment and control study designs. It combines two desirable properties: it is designed for multivariate settings while at the same time capturing the heterogeneity in treatment response of individuals. In particular, it complements both the classical difference-in-differences estimator, which is applicable in multivariate settings but only captures average effects, and the changes-in-changes estimator \citep{athey2006identification}, which allows for general treatment heterogeneity but is only applicable in univariate settings.

Showcasing the utility of the proposed method, we revisit the classical Card \& Krueger dataset by decomposing the treatment effects for full-and part-time employees. We find that fast-food restaurants responded to an increased minimum wage by substituting full-time employees for part-time employees, even when accounting for restaurant size. This provides a novel insight by dissecting the relationship between full- and part-time employees while accounting for the heterogeneity in restaurant size. More generally, since our proposed method is able to consider multivariate outcomes, it can help uncover interesting causal effects and nonlinear relations in a wide array of applications.

\section{Authors' Statements}

\subsection{Funding Information}
William Torous was supported as an undergraduate through MIT's UROP program with funding from the Lord Foundation and the Paul E. Gray UROP Fund. Florian Gunsilius was supported by a MITRE faculty research award. Philippe Rigollet was supported by NSF grants IIS-1838071, DMS-2022448, and CCF-2106377.

\subsection{Conflict of interest}
The authors state no conflict of interest.

\subsection{Data Availability Statement}
The datasets analyzed during the current study are available at David Card's public website, \url{https://davidcard.berkeley.edu/data_sets.html}. 

Implementations of our methodology in $\texttt{R}$,  $\texttt{Python}$, and $\texttt{STATA}$ are under development.

\bibliography{ref}

\newpage
\appendix

\section{Proofs of Stated Results}\label{sec:proofs}

For compactness and clarity we use a different notation for measures in these proofs than in the main text. For control group outcomes, $\Prob_{Y_{C,0}}$ is equivalent here to $\mu_0$ and $\Prob_{Y_{C,1}}$ is equivalent to $\mu_1$. For the treated, $\Prob_{Y_{T,0}}$ is equivalent to $\mu_0^\star$ and $\Prob_{Y_{T,1}}$ is equivalent to $\mu_1^\star$. The counterfactual measure $\Prob_{Y_{T,1}^\dagger}$ has $\mu_1^\dagger$ as its counterpart.

\subsection{Proof of Theorem \ref{thm:main}}\label{sec:proof_main}
\begin{proof}
The cyclically comonotonicity assumption reads as 
\[\sum_{i=1}^m\langle h_0(u_{i}), h_1(u_{i+1})- h_1(u_{i})\rangle \le 0\] for all $m\in\mathbb{N}$.
Define $y_i=h_0(u_i)$ and $y_{i+1}=h_0(u_{i+1})$ so that 
\[
\sum_{i=1}^m \langle y_{i},h_1\circ h_0^{-1} (y_{i+1})-h_1\circ h_0^{-1} (y_{i})\rangle\leq 0
\]
for all $x_0, y_0$ in the image of $h_0$. This means that the function $\sd=h_1 \circ h_0^{-1}$ is cyclically monotone in the sense of ~\eqref{eq:CM}. By \cite[Theorem 24.8]{rockafellar1997convex}, this is equivalent to $\sd$ being the gradient of a convex function and hence the unique Brenier map from $\mu_0$ to $\mu_1$. Moreover, by Assumption \ref{ass:measures} $K_0^*\subset K_0$, so that we can apply the map $\sd$ to $\mu_0^\star$ to identify $\mu_1^\dagger$.
\end{proof}

\subsection{Proof of Proposition \ref{prop:consistency}}\label{sec:proof_consistency}

\begin{proof}
The main result to enable a consistency statement for $\hat{\mu}_{1,m}^\dagger$ will be the uniform convergence for optimal maps presented in Theorem 1.1 of \citet{segers2022Graphical}. This states that with the Euclidean norm

$$ \lim\limits_{n \to \infty} \sup_{x_0 \in K_0} \sup_{y \in \sdhat_{\# \{x_0\} }} \lVert y - \sd_{\# \{ x_0 \} } \rVert \to 0.$$

In other words, the pushforward of singletons under $\sdhat$ converges uniformly to their population pushforward under $\sd$. The above statement holds by case (b) of that theorem, and its assumptions are justified as follows. The empirical source measure $\hat{\mu}_{0,n}$ and target measure $\hat{\mu}_{1,l}$ both converge weakly in probability to their population distributions \citep{varadarajan1958Weak}. The Brenier map $\sd$ is the unique coupling with cyclically monotone support between $\mu_0$ and $\mu_1$. Finally, both $\sdhat$ and $\sd$ are maximally cyclically monotone functions on the non-empty supports of their respective coupling measures by results in \citet{rockafellar1966characterization}.

For finite $n$, the pre-image of $\sdhat$ will not contain all possible singletons in $K_0^\star \subset K_0$. Instead we generate counterfactuals by rounding treatment samples $x_0^\star$ to their nearest neighbor in the control pre-image, $\tilde{x}_0^\star$, and using $\sdhat_{\# \{ \tilde{x}_0^\star \} }$ as the counterfactual estimator. We wish to show

$$ \lim\limits_{n \to \infty} \sup_{x_0^\star \in K_0^\star} \sup_{y \in \sdhat_{\# \{ \tilde{x}_0^\star \}} } \lVert y - \sd_{\# \{x_0^\star \} } \rVert \to 0.$$

The exchange of $x_0^\star$ for $x_0$ in the above statement follows because the set $\mathcal{C}$ of all closed balls on $\mathbb{R}^d$ is a Vapnik–Chervonenkis class with VC dimension $d+2$ \citep[pg.~833]{shorack2009empirical, vapnik1971uniform}.  Therefore, with $\sigma(\cdot)$ indicating the Borel $\sigma$-algebra and $\|\cdot\|_{\mathcal{C}}$ denoting the supremum of the Euclidean distance evaluated over the set of closed balls,
$$\sup_{\mu_0 \in \mathcal{P}(K_0, \sigma(K_0))} \E \lVert \hat{\mu}_{0,n} - \mu_0 \rVert_\mathcal{C} \to 0, \ n \to \infty.$$

This implies for sufficiently large $n$ the empirical $\hat{\mu}_{0,n}$ will contain a sample from all closed balls. It follows that for any $\varepsilon > 0$, $\tilde{x}_0^\star$, the nearest neighbor of $x_0^\star$, must be an element of $B_\varepsilon(x_0^\star)$. 

We now consider
\begin{multline*}
\limsup_{\min(m,n) \to \infty} \int_{K_0^\star} \lVert \sdhat(\tilde{x}) - \sd(x) \rVert \mu_0^\star(dx) \le\\ \limsup_{\min(m,n) \to \infty} \left[ \int_{K_0^\star} \lVert \sdhat (\tilde{x}) - \sd(\tilde{x}) \rVert  \mu_0^\star(dx) + \int_{K_0^\star} \lVert \sd(\tilde{x}) - \sd(x) \rVert  \mu_0^\star(dx) \right].
\end{multline*}

The first integral must converge to $0$ in the limit supremum because of the uniform convergence for optimal maps stated above. The second integral involves a Brenier map $\sd$ from $\mu_0^\star$ to $\mu_1^\dagger$ which is $\mu_0^\star$-Lipschitz almost everywhere. This implies $\int_{K_0^\star} \lVert \sd(\tilde{x}) - \sd(x) \rVert  \mu_0^\star(dx) \le C \int_{K_0^\star} \lVert \tilde{x} - x \rVert  \mu_0^\star(dx) \le C \sup_{K_0^\star} \lVert \tilde{x} - x \rVert$ for some Lipschitz constant $C<+\infty$. We have already shown that in a sufficiently large sample, all nearest neighbor matches $\tilde{x}$ will be arbitrarily close to $x$. Therefore, the limit supremum of this integral is bounded by $C \varepsilon$ for any $\varepsilon > 0.$ It follows that $$\limsup_{\min(m,n) \to \infty} \int_{K_0^\star} \lVert \sdhat(\tilde{x}) - \sd(x) \rVert \mu_0^\star(dx) = 0.$$

This integral also upper bounds $\lvert \int_{K_0^\star} \lVert \sdhat (\tilde{x}) - x \rVert  \mu_0^\star(dx) -  \int_{K_0^\star} \lVert \sd (x) - x \rVert  \mu_0^\star(dx) \rvert $ by the reverse triangle inequality, so this absolute difference must also converge to zero in the $\limsup$. 
The sequence of empirical $\hat{\mu}_{0,m}^\star$ converge weakly in probability to $\mu_0^\star$. Segers' theorem implies uniform convergence, so
\[\int_{K_0^\star} \lVert \sdhat (\tilde{x}) - x \rVert  \hat{\mu}_{0,m}^\star(dx)\leq C \sup \|\sdhat (\tilde{x}) - x\| \cdot \hat{\mu}_{0,m}^\star(K_0^\star) \to 0.\] H\"older's inequality and the triangle inequality gives

\begin{align*}
&\left\lvert \int_{K_0^\star} \lVert \sdhat (\tilde{x}) - x \rVert  \hat{\mu}_{0,m}^\star(dx) - \int_{K_0^\star} \lVert \sdhat (\tilde{x}) - x \rVert  \mu_0^\star(dx)\right\rvert\\
\leq &\left\lvert \int_{K_0^\star} \lVert \sdhat (\tilde{x}) - x \rVert  \left[\hat{\mu}_{0,m}^\star -\mu_{0}^\star\right](dx)\right\rvert +\left\lvert \int_{K_0^\star} \lVert \sdhat (\tilde{x}) - x \rVert  \mu_{0}^\star(dx) - \int_{K_0^\star} \lVert \sdhat (\tilde{x}) - x \rVert  \mu_0^\star(dx)\right\rvert\\
\leq & \sup \|\sdhat (\tilde{x}) - x\| \cdot \left[\hat{\mu}_{0,m}^\star -\mu_{0}^\star\right](K_0^\star) + \sup \|\sdhat (\tilde{x}) - x\| \cdot \mu_{0}^\star(K_0^\star) \to 0
\end{align*}
so we have convergence in probability of $\int_{K_0^\star} \lVert \sdhat (\tilde{x}) - x \rVert  \hat{\mu}_{0,m}^\star(dx)$ to $\int_{K_0^\star} \lVert \sdhat (\tilde{x}) - x \rVert  \mu_0^\star(dx)$. Combining these results, we can state that for all $\varepsilon > 0$, $${\lim\limits_{\min(m,n)} \mathbb{P} \left( \Big\lvert \int_{K_0^\star} \lVert \sdhat (\tilde{x}) - x \rVert  \hat{\mu}_{0,m}^\star(dx) - \int_{K_0^\star} \lVert \sd (x) - x \rVert  \mu_0^\star(dx) \Big\rvert > \varepsilon \right) \to 0}.$$
\end{proof}

\subsection{Proof of Proposition \ref{prop:rate}}\label{sec:proof_rate}

\begin{proof}
We wish to bound the risk between the true and estimated counterfactual outcome distributions with respect to the 2-Wasserstein metric: $\E W_2 (\sd_{\# \mu_0^\star}, \sdhat_{\# \tilde{\mu}_{0,m}^\star})$. We will show this quantity is $O(n^{\nicefrac{-1}{d}})$. Note that this expectation is jointly over $\mu_0^\star$, control measure $\mu_0$ onto whose sampled support we project, and $\mu_1$ through estimating $\sdhat$. The assumed mutual independence between all samples will sometimes allow us to marginalize out distributions from an expectation; the distributions which we jointly take expectations over will be noted in that operator's subscript. 

Applying the triangle inequality we have
\begin{align} \label{eq:main_traingle}
    \E_{\mu_0,\mu_1,\mu_0^\star} W_2 (\sd_{\# \mu_0^\star}, \sdhat_{\# \tilde{\mu}_{0,m}^\star}) \le 
\E_{\mu_0, \mu_0^\star} W_2 (\sd_{\# \mu_0^\star}, \sd_{\# \tilde{\mu}_{0,m}^\star}) + \E_{\mu_0, \mu_0^\star, \mu_1} W_2 (\sd_{\# \tilde{\mu}_{0,m}^\star}, \sdhat_{\# \tilde{\mu}_{0,m}^\star}).
\end{align}

Consider first $\E_{\mu_0, \mu_0^\star} W_2 (\sd_{\# \mu_0^\star}, \sd_{\# \tilde{\mu}_{0,m}^\star})$. We will bound the expected Wasserstein distance from $\mu_0^\star$ to its projection on the controls' support $\tilde{\mu}_{0,m}^\star$. Then we will show this convergence rate does not change under pushforward of both measures by $\sd$.

We have $\E_{\mu_0, \mu_0^\star} W_2(\mu_0^\star, \tilde{\mu}_{0,m}^\star) \le \E_{\mu_0} W_2(\mu_0^\star, \hat{\mu}_{0,m}^\star) + \E_{\mu_0, \mu_0^\star} W_2(\hat{\mu}_{0,m}^\star, \tilde{\mu}_{0,m}^\star)$ again by the triangle inequality. The first term $\E_{\mu_0^\star} W_2(\mu_0^\star, \hat{\mu}_{0,m}^\star)$ is the expected distance between the true distribution and its i.i.d sample; we have an $O(m^{\nicefrac{-1}{d}})$ upper-bound e.g. from \citet{fournier2015rate}. The second term is the expected Wasserstein distance between sample $\hat{\mu}_{0,m}^\star$ and its nearest neighbor match in the support of $\hat{\mu}_{0,n}$. 

The distributions of these empirical measures for a fixed sample size are denoted  $\sigma_{0,n}$, $\sigma_{0,m}^\star$, and $\sigma_{1,l}$, respectively. We can express the joint expectation $\E_{\mu_0, \mu_0^\star} W_2(\hat{\mu}_{0,m}^\star, \tilde{\mu}_{0,m}^\star)$ as the marginalized integral over the treatment group's sampling distribution $\int_{\hat{\mu}_{0,m}^\star} \E_{\mu_0} W_2(\hat{\mu}_{0,m}^\star, \tilde{\mu}_{0,m}^\star) \sigma_{0,m}^\star(\hat{\mu}_{0,m}^\star)$. We wish show that for fixed 
$\hat{\mu}_{0,m}^\star$, $\E_{\mu_0} W_2(\hat{\mu}_{0,m}^\star, \tilde{\mu}_{0,m}^\star) = O(n^{\nicefrac{-1}{d}})$ which will allow us to conclude $\E_{\mu_0, \mu_0^\star} W_2(\hat{\mu}_{0,m}^\star, \tilde{\mu}_{0,m}^\star) = O(n^{\nicefrac{-1}{d}})$. 

Lemma 3.1 from \citet{canas2012Learning} states $W_2^2(\mu, \pi_S \mu) = \E_{\mu} \lVert x - S \rVert^2$ where $\pi_S$ is the nearest neighbor projection onto set $S$.
In words, squared Euclidean nearest neighbor matching solves an optimal transport problem. Let $S = \text{supp} \ \hat{\mu}_{0,n}$.
This formulation allows us to express $\int_{\hat{\mu}_{0,m}^\star} \E_{\mu_0} W_2^2(\hat{\mu}_{0,m}^\star, \tilde{\mu}_{0,m}^\star) \sigma_{0,m}^\star(\hat{\mu}_{0,m}^\star)$ as $\int_{\hat{\mu}_{0,m}^\star} \E_{\mu_0} \E_{\hat{\mu}_{0,m}^\star} \lVert x - S \rVert^2 \sigma_{0,m}^\star(\hat{\mu}_{0,m}^\star)$. Replacing $\E_{\hat{\mu}_{0,m}^\star}$ with a summation yields \\ $\int_{\hat{\mu}_{0,m}^\star} \nicefrac{1}{m} \sum_{i=1}^m \E_{\mu_0} \lVert x_i - S \rVert^2 \sigma_{0,m}^\star(\hat{\mu}_{0,m}^\star)$. 
Lemma 1 of \citet{abadie2006Large} provides bounds on the matching discrepancy between fixed point $z$ and potential matches drawn i.i.d from a distribution with compact, convex, and bounded support. Denoting the discrepancy vector to the nearest neighbor $U_1$, they find $\E_{\mu_0} \lVert U_1 \rVert ^3 = O(n^{-\nicefrac{3}{d}})$.
Applying the Lyapunov's inequality yields $\E_{\mu_0} \lVert U_1 \rVert ^2 = O(n^{-\nicefrac{2}{d}})$. Thus have $\E_{\mu_0} W_2^2(\hat{\mu}_{0,m}^\star, \tilde{\mu}_{0,m}^\star) = O(n^{-\nicefrac{2}{d}})$. From Jensen's $\left( \E_{\mu_0} W_2^2(\hat{\mu}_{0,m}^\star, \tilde{\mu}_{0,m}^\star) \right)^{\nicefrac{1}{2}} \geq \E_{\mu_0} W_2(\hat{\mu}_{0,m}^\star, \tilde{\mu}_{0,m}^\star) = O(n^{\nicefrac{-1}{d}})$. 
Integrating this bound over the sampling distribution does not change it. Therefore, we can conclude by summing the two triangle inequality rates to yield $\E_{\mu_0, \mu_0^\star} W_2(\hat{\mu}_{0,m}^\star, \tilde{\mu}_{0,m}^\star) = O(\text{min}(m,n)^{\nicefrac{1}{d}})$. 
Results to shortly follow will show this rate holds for the pushforward $\E_{\mu_0, \mu_0^\star} W_2 (\sd_{\# \mu_0^\star}, \sd_{\# \tilde{\mu}_{0,m}^\star})$ as well. 

Now we turn our attention to the rate of convergence for $\E_{\mu_0, \mu_0^\star, \mu_1} W_2 (\sd_{\# \tilde{\mu}_{0,m}^\star}, \sdhat_{\# \tilde{\mu}_{0,m}^\star})$. We once again appeal to the triangle inequality to upper bound this expectation. Note that

\begin{align} \label{eq:sub_traingle}
   \E_{\mu_0, \mu_0^\star, \mu_1} W_2 (\sd_{\# \tilde{\mu}_{0,m}^\star}, \sdhat_{\# \tilde{\mu}_{0,m}^\star}) \leq \notag\\ \E_{\mu_0, \mu_0^\star} W_2 (\sd_{\# \tilde{\mu}_{0,m}^\star}, \sd_{\# \mu_0^\star}) + \E_{\mu_0, \mu_0^\star, \mu_1} W_2(\sd_{\# \mu_0^\star}, \sdhat_{\# \mu_0^\star}) + \E W_2(\sdhat_{\# \mu_0^\star}, \sdhat_{\# \tilde{\mu}_{0,m}^\star}).
\end{align}

To show the first and third terms are $O(\text{min}(m,n)^{\nicefrac{1}{d}})$, it is sufficient to show the rate of converge between two measures does not change under pushforward via an optimal transport map. Recall that optimal transport maps are twice differentiable almost everywhere, which implies they are also Lipschitz almost everywhere. 

Let $\mathsf{r}$ denote the rounding optimal transport map from $\mu_0^\star$ to $\tilde{\mu}_{0,m}^\star$. For all $x$ in the support $\mu_0^\star$ we have ${\lVert \sd \circ \mathsf{r} (x) - \sd(x) \rVert \le K \lVert \mathsf{r}(x) - x \rVert}$ for some Lipschitz constant $K$. Integrating both sides over $\mu_0^\star$ preserves this inequality and yields $W_2(\sd_{\# \tilde{\mu}_{0,m}^\star}, \sd_{\# \mu_0^\star}) \le K \ W_2(\tilde{\mu}_{0,m}^\star, \mu_0^\star)$. It follows $\E_{\mu_0, \mu_0^\star}  W_2(\sd_{\# \tilde{\mu}_{0,m}^\star}, \sd_{\# \mu_0^\star}) = O(\text{min}(m,n)^{\nicefrac{1}{d}})$. Adapting this argument with $\sdhat$ instead of $\sd$ shows that $\E_{\mu_0, \mu_0^\star, \mu_1}  W_2(\sdhat_{\# \tilde{\mu}_{0}^\star}, \sdhat_{\# \tilde{\mu}_{0,m}^\star}) = O(\text{min}(m,n)^{\nicefrac{1}{d}})$ as well.

It remains to show $\E_{\mu_0, \mu_0^\star, \mu_1} W_2(\sd_{\# \mu_0^\star}, \sdhat_{\# \mu_0^\star}) = O(n^{\nicefrac{-1}{d}})$. The Radon-Nikodym derivative $f$ for $\mu_0^\star$ with respect to $\mu_0$ exists and is bounded by assumption. From the definition of a push forward, we see that $W_2(\sd_{\# \mu_0^\star}, \sdhat_{\# \mu_0^\star})$ can be expressed as $\int_{K_0} \lVert \sd(x) - \sdhat(x) \rVert f(x) \mu_0(dx)$. It is well known that $\int_{K_0} \lVert \sd(x) - \sdhat(x) \rVert \mu_0(dx) = O(n^{\nicefrac{-1}{d}})$; see e.g. \citet{manole2021sharp}. Let $C$ equal the upper bound of $f$. Clearly $\int_{K_0} \lVert \sd(x) - \sdhat(x) \rVert f(x) \mu_0(dx) \le C \int_{K_0} \lVert \sd(x) - \sdhat(x) \rVert \mu_0(dx) = O(n^{\nicefrac{-1}{d}})$.

We have shown each term of Equation \ref{eq:sub_traingle} converges at a rate of $O(\text{min}(m,n)^{\nicefrac{1}{d}})$. Therefore, \linebreak $\E_{\mu_0, \mu_0^\star, \mu_1} W_2 (\sd_{\# \tilde{\mu}_{0,m}^\star}, \sdhat_{\# \tilde{\mu}_{0,m}^\star}) = O(\text{min}(m,n)^{\nicefrac{1}{d}})$. Both terms on the right-hand side of Equation \ref{eq:main_traingle} converges at a rate of $O(\text{min}(m,n)^{\nicefrac{1}{d}})$, so $\E_{\mu_0, \mu_0^\star, \mu_1} W_2({\mu_1^\dagger, \hat{\mu}_{1,m}^\dagger) = O(\text{min}(m,n)^{\nicefrac{1}{d}})}$ as desired.
\end{proof}

\section{Calculating Optimal Transport Couplings from Data}\label{sec:estimation_from_data}

This section outlines how to estimate optimal transport couplings from sampled data and provides references to supporting theory. Given $n$ samples from measure $\mu$ and $m$ samples from measure $\nu$, both supported on $\R^d$, the objective is to estimate the Brenier map $\bar T$ such that $\nu = \bar T_\# \mu$. The Brenier map's definition is given by ~\eqref{eq:OT}.

In practice, the measures $\mu$ and $\nu$ are replaced by their empirical analogues, denoted $\hat \mu\coloneqq\frac{1}{n}\sum_{i=1}^n \delta_{X_i}$ and $\hat \nu \coloneqq\frac{1}{m}\sum_{j=1}^m \delta_{Y_j}$, which are discrete uniform distributions over their samples.\footnote{For any measurable set $A$, $\delta_x(A)$ denotes the Dirac measure which is $1$ if $x\in A$ and $0$ otherwise.} The transport cost between observations can be summarized by the matrix $\textbf{C} \in \R_{\geq 0}^{n \times m}$ where $\textbf{C}_{i, j} = c(\hat \mu[i],\hat \nu[j])$, which is defined with respect to a cost function $c(x,y): \R^d \times \R^d \mapsto [0, \infty)$. Theorem \ref{thm:brenier} holds under the squared Euclidean distance: $c(x,y) = \|x-y\|^2$, which corresponds to the 2-Wasserstein distance between measures. A non-negative transport matrix $\textbf{T}$ which moves the atoms of $\hat \mu$ to $\hat \nu$ is the object to optimize over. We enforce the recovery of marginals in the following discrete analog of ~\eqref{eq:OT}'s optimization:

\begin{equation}
    \label{eq:discete_OT}
    \min_{\mathbf{T} \in \R_{\geq 0}^{n \times m}}  \sum_{i, j} \textbf{T}_{i,j} \textbf{C}_{i,j} \qquad \text{ s.t. }\thickspace n\textbf{T}\mathds{1}_m = \mathds{1}_n \quad\text{ and }\quad m\textbf{T}^\intercal\mathds{1}_n = \mathds{1}_m.
\end{equation}

Here $\mathds{1}_n$ denotes the length $n$ vector with all entries equal to $1$. Noting that ~\eqref{eq:discete_OT} can be reformulated as a linear program, it is common to apply algorithms from that domain which yield bijective maps when $n=m$ and sparse plans otherwise. Another approach is to add an entropic penalty term to ~\eqref{eq:discete_OT} which relaxes the problem and enables computation through Sinkhorn's algorithm \citep{cuturi2013sinkhorn, peyre2019computational}. Sinkhorn-based approaches add the penalty term $\lambda \mathbf{H}(\mathbf{T})$ to ~\eqref{eq:discete_OT}, where $\mathbf{H}(\mathbf{T})$ is the matrix entropy of $\mathbf{T}$ and positive tuning parameter $\lambda$ controls the sparsity of $\mathbf{T}$. For large choices of $\lambda$, the optimal solution will be driven towards the naive coupling given by the outer-product of marginals; for small $\lambda$, sparse solutions more akin to the unregularized problem are found. Without loss of generality assuming $n \geq m$, the runtime of linear programming approaches is generally $O(n^3 \log(n))$ while Sinkhorn can achieve $O(n^2 \varepsilon^{-3})$ if we wish to bound the error of our discretized Wasserstein distance by $\varepsilon$ \citep{altschuler2017near}.

Our methodology involves first estimating $\hat \sd$ from control group observations and then using that estimate to pushforward $\mu_0^\star$. Representing $\hat \sd$ as a matrix, we can recover its pushforward of $\hat \mu_0$ through $n\textbf{T} \hat \mu_1$. As it is unlikely that the supports of the empirical measures $\hat \mu_0$ and $\hat \mu_0^\star$ have a large intersection, we must extrapolate $\hat \sd$'s behavior on $\hat \mu_0^\star$. A common approach is using nearest neighbors \citep{papadakis2015image}. Counterfactual treatment outcomes are generated by matching each treatment unit pre-intervention to the nearest control and then pushing forward along the transport plan between observed control distributions.
Neural network-based approaches to learn maps have also been proposed \citep{seguy2018largescale, makkuva2020optimal}, and representation results help support this direction of active research \citep{lu2020universal}.

To compute optimal transport maps in this paper, we use a linear programming method \texttt{ot.lp.emd} from the Python package \texttt{Python Optimal Transport} \citep{flamary2021pot}. These maps are then extrapolated into optimal transport plans with Euclidean nearest neighbor matching. In our simulations the control and treatment groups pre-intervention have almost complete overlap of sampled support, justifying this technique. All numerical experiments in this paper are written and run in Python 3. Replication code is available in the supplementary materials for this paper.

\section{Numerical Experiments}\label{sec:more_simulations}

In this section we provide implementation details for results computed in Sections \ref{sec:motivating_example} and \ref{sec:numerical_experiment}. 

\subsection{Implementation Details for Numerical Experiment in Section \ref{sec:motivating_example}}\label{sec:motivating_example_details}

We consider $n = 3000$ units in $\mathbb{R}^2$ for each treatment arm. Samples of latent variables are drawn from the distribution $\nu$ at $t=0$ and $t=1$, mirroring the setup discussed in Section \ref{sec:natural_drift}. Draws from $\nu^\star$ are fixed across time for counterfactual validation purposes. For controls, $\nu$'s first coordinate is distributed according to ${\sf Beta}(3,2)$ and second to ${\sf Beta} (2,3)$; for treatments, $\nu^\star$'s first coordinate is distributed according to ${\sf Beta} (2,3)$ and second to ${\sf Beta} (3,2)$. 

Recall that we consider the following set of production functions that are cyclically comonotone:
\[h_0(u) = \begin{bmatrix} 1&\alpha\\\alpha& 1\end{bmatrix}u\qquad\text{and}\qquad h_1(u) = \begin{bmatrix} 1&-\alpha\\-\alpha& 1\end{bmatrix}u.\] 

In our simulations, the $\alpha$ parameter of the production functions is fixed at $0.5$. The proof of their cyclically comonotone relationship when $|\alpha| \in (0,1)$ follows.
\begin{proof}
We begin by simplifying the inner product terms contained in the definition of cyclical comonotonicity ~\eqref{eq:CCM}. We have
$$
\langle h_0(u_i), h_1(u_i) - h_1(u_{i+1}) \rangle = \Big \langle 
\begin{bmatrix} 1&\alpha\\\alpha& 1\end{bmatrix}u_i,
\begin{bmatrix} 1&-\alpha\\ -\alpha& 1\end{bmatrix}(u_i - u_{i+1})
\Big \rangle.
$$
After expansion and simplification, this reduces to $(1 - \alpha)^2 \langle u_i, u_i - u_{i+1} \rangle$. Substituting this into the definition of ~\eqref{eq:CCM}, we now wish to show 
$(1 - \alpha)^2 \sum_{i=1}^m\langle u_{i}, u_{i} - u_{i+1})\rangle \geq 0$ for all $m$. Our restriction of $|\alpha| \in (0,1)$ ensures $(1 - \alpha)^2$ is always positive, and this summation is equivalent to the identity map being cyclically monotone in the sense of ~\eqref{eq:CM}. The identity map is the gradient of a convex function, $0.5 ||u||_2^2$, completing our proof.
\end{proof}

Using these production functions, we generate $n$ independent samples from each of the distributions $\mu_0, \mu_1$ and $\mu_0^\star$, as well as samples from the true counterfactual distribution $\mu_1^\dagger$ for validation purposes.

To estimate the transport map $\sd$, we first compute an optimal plan using observed data and round it to an optimal transport map. This map is only defined on the data from $\mu_0$. To predict counterfactuals treatment on data from $\mu_0^\star$, we employ nearest neighbor interpolation. Since the Beta distribution is supported on $[0,1]$ for all parameter choices, $\mu_0$ and $\mu_0^\star$ have identical support in this example and this naive extrapolation technique performs sufficiently well. In particular, it shows that the OT based estimator remains close to the true counterfactual distribution while naive tensorization of CiC may depart significantly.

\begin{figure}[ht!]
    \centering
    \begin{minipage}[b]{0.4\textwidth}
    \includegraphics[width=\textwidth]{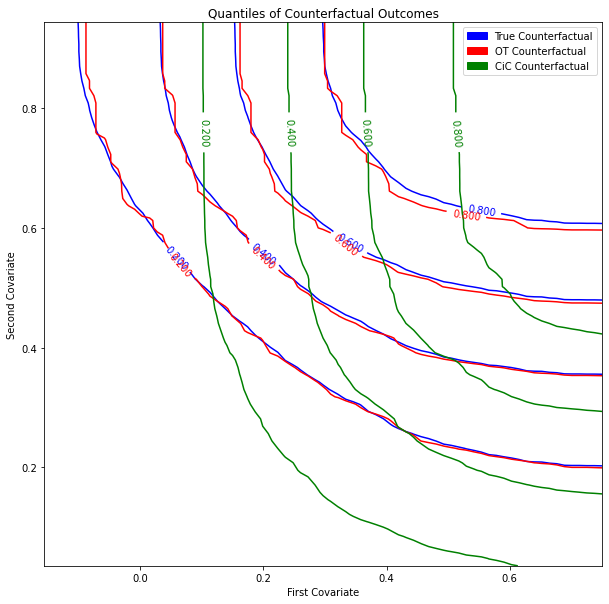}
    \caption{Quantiles of the empirical CDF for the bivaraite experiment over a uniform mesh with $10,000$ points.}
    \label{fig:ecdf_plot}
    \end{minipage}
\end{figure}

We also compute the empirical CDF for our $3000$ true counterfactual observations and the counterfactuals generated by the two methods over a uniform mesh of $10,000$ points. Figure \ref{fig:ecdf_plot} visually demonstrates that OT almost perfectly recovers the eCDF, and Table \ref{table:experimental_results} quantifies this result. The mean absolute difference over each mesh point is an order of magnitude smaller for OT and has a smaller standard deviation. 

\begin{table}[h]
\centering
\caption{Recovery of eCDF by Method Over 20 Runs.}\label{table:experimental_results}
\begin{tabular}{c|c|c}
&MAD OT&MAD CiC\\\hline\hline
Mean&$.008$&$.089$\\\hline
st. dev.&$.002$&$.003$\\\hline\hline
\end{tabular}\vspace{.3cm}
\end{table}

\subsection{Implementation Details for Numerical Experiment in Section
\ref{sec:numerical_experiment}}\label{sec:numerical_experiment_details}

Recall that we consider a class convex functions indexed by $\theta$ where $g(x; \theta) = \sum_{i=1}^p x_{(i_1)}^2  x_{(i_2)}^{-1}$. Here $x \in \mathbb{R}_{++}^d$, $x_{(i)}$ represents the $i^{\text{th}}$ coordinate of $x$, $i_j \in [[d]]$ for $j \in \{1, 2\}$, and the parameter $\theta$ contains the $p$ pairs of indices to be compared. 

It is well known the that function $f = x^2/y$ is convex when $y >0$. Furthermore, the sum of convex functions remains convex. Therefore, the functions $g(x; \theta)$ are convex. In our numerical experiment, we sample $\theta$ randomly in each run to contain $d$ coordinate-pair interactions, matching the dimension of the latent vectors. 

In our experiment we define $h_0$ as the identity and $h_1 = \nabla g$. Therefore, cyclical comonotonicity ~\eqref{eq:CCM} is equivalent to the cyclical monotonicity ~\eqref{eq:CM} of $g$. As $g$ is convex, $\nabla g$ is an optimal transport map, and therefore cyclically monotone. 

The $d$ coordinate pairs randomly selected from a uniform distribution on $[1, d]$. As these results only hold for strictly positive latent vectors, we sample both latent distributions from ${\sf Beta}(2,3)$. Both treatment arms have $n=3,000$ units.

The implementation of our experiment is otherwise identical to that described in Appendix \ref{sec:motivating_example_details}; we generate data for the four relevant distributions and compute transport plans with nearest neighbor extrapolation. The eCDF and MAD computation is also the same.

\section{Additional Card \& Krueger Experiment}\label{sec:ck_appendix}

\begin{figure}[h]
    \centering
    \begin{minipage}[b]{0.5\textwidth}
    \centering
    \includegraphics[width=\textwidth]{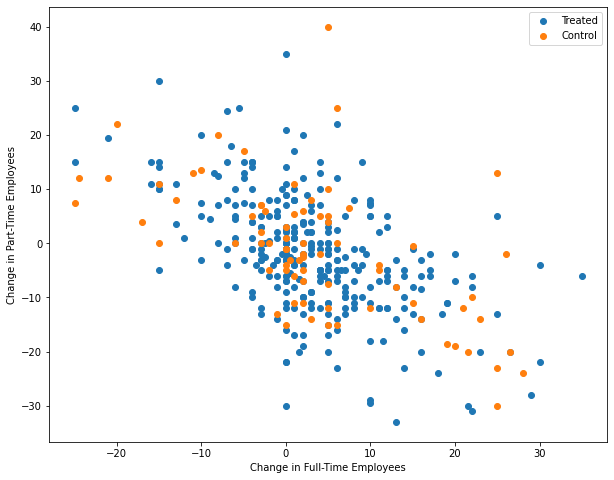}
    \caption{Distribution of unit-level treatment effect estimates, demonstrating a negative correlation between change in full-time and part-time employees.}
    \label{fig:CK_scatter}
    \end{minipage}
\end{figure}

We begin by presenting an additional figure from our bivariate Card \& Krueger analysis in Section \ref{sec:ck_reanalysis}. Figure \ref{fig:CK_scatter} demonstrates a negative correlation between the change in full-time employees after the intervention and the change in part-time employees. This suggests that fast food restaurants substituted full-time employees for part-time ones.

\begin{figure}[ht!]
    \centering
     \begin{minipage}[b]{0.45\textwidth}
         \centering
         \includegraphics[width=\textwidth]{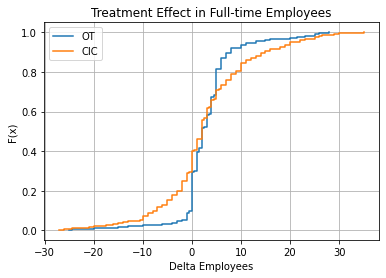}\\\vspace{.2cm}
         (a) Full-time Employment
         \label{fig:ck_first_marginal}
     \end{minipage}
     \begin{minipage}[b]{0.45\textwidth}
         \centering
         \includegraphics[width=\textwidth]{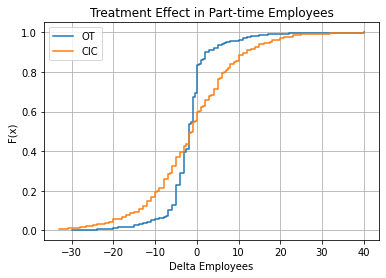}\\\vspace{.2cm}
         (b) Part-time Employment
         \label{fig:ck_second_marginal}
     \end{minipage}
     \hfill
        \caption{Recovery of counterfactual marginals by OT and CiC.}
        \label{fig:ck_marginals}
\end{figure}  

For our reanalysis of the Card \& Krueger data on the minimum wage and fast food employment, we use the original dataset provided by David Card on his website. It contains survey information, collected via telephone, for $410$ fast food restaurants across $4$ fast-food chains. We remove $19$ restaurants with missing outcome observations. Of the remaining, $76$ are in the control region of northeastern Pennsylvania and $315$ are in New Jersey. In addition to the two employment outcomes of interest, a number of other observations about a restaurant's wages, prices, and operations were collected. Our optimal transport methodology allows us to estimate treatment effects more richly with potential outcomes including non-employment time-varying restaurant characteristics. We emphasize that this experiment is presented to illustrate the applicability of our method in higher dimensions. Our variable selection is not data-driven and the ratio of covariate to units is low.

\begin{table}[ht!]
\centering
\caption{Outcomes and Covariates Included in Our Analysis.}\label{table:ck_covariates}
\scalebox{0.75}{
\begin{tabular}{c|c}
Covariate Name &Description\\\hline\hline
\texttt{EMPFT}&Number of full-time employees\\\hline
\texttt{EMPPT}&Number of part-time employees\\\hline\hline
\texttt{PCTAFF}&Percent of employees affected by new minimum wage\\\hline
\texttt{NMGRS}&Number of managers\\\hline
\texttt{INCTIME}&Months until usual first raise\\\hline
\texttt{PENTREE}&Price of an entrée with tax\\\hline
\texttt{PSODA}&Price of a soda with tax\\\hline
\texttt{NREGS}&Number of cash registers in the restaurant\\\hline
\texttt{OPEN}&Hour of opening\\\hline
\texttt{HRSOPEN}&Number of hours open per day\\\hline\hline
\end{tabular}}\vspace{.3cm}
\end{table}

We select a representative subset of $10$ numerical covariates, listed in Table \ref{table:ck_covariates}, and remove any restaurant with a missing entry for any of these covariates. This leaves a final sample of $52$ controls and $200$ treatments. The full dataset has $15$ numerical covariates but including all $15$ leads to a small sample size due to missing data. The covariates we do not include add little additional information, such as the price of fries (we include the price of entrées and sodas), have high rates of missingness, or have a different distribution in the subsample with complete data than the entire survey population. Unlike the outcome-only analysis presented in Section \ref{sec:ck_reanalysis}, the included covariates have different scales. Thus, we standardize each to zero mean and unit standard deviation.  The dataset also includes categorical data such as the fast-food chain, which we do not include because the Euclidean distance between restaurants would depend on how these are encoded. 

Instead of presenting results conditional on some set of covariates, we present aggregate results over subsets of all subsets of potential outcome vectors including both employment outcomes and 2, 3, or 4 other covariates. The positive results of this experiment are included in Table \ref{table:robustness_results}. The sign of our two estimates never change in this experiment. Furthermore, addings covariates leads to optimal transport estimators with smaller estimated full-time ATE and larger part-time ATE on aggregate. In future work with more tests for covariate selection and further sensitivity analysis, we hope stronger claims about the estimates' magnitude can be made.
\begin{table}[t!]
\centering
\caption{Summary Statistics of Higher Dimensional Subsets for the Card \& Krueger ATE.}
\begin{tabular}{c|c|c}
&TE FT&TE PT\\\hline\hline
Mean&$1.56$&$-1.97$\\\hline
st. dev.&$0.38$&$0.56$\\\hline
Min&$0.69$&$-3.59$\\\hline
Max&$2.63$&$-0.78$\\\hline\hline
\end{tabular}
\label{table:covar_robustness}
\label{tab:unit_robustness}

\label{table:robustness_results}
\end{table}

\end{document}